\documentclass[11pt,a4paper]{article}
\usepackage{microtype}
\usepackage{lmodern}
\usepackage[T1]{fontenc}
\usepackage{color}

\usepackage{dcolumn}
\usepackage{bm}
\usepackage{mathrsfs}
\usepackage[T1]{fontenc}
\usepackage{cite}
\usepackage{jheppub}
\usepackage{amsmath}
\usepackage{rotating}
\usepackage{slashed}

\def\beq{\begin{equation}}
\def\eeq{\end{equation}}
\newcommand{\beqa}{\begin{eqnarray}} 
\newcommand{\eeqa}{\end{eqnarray}}
\newcommand{\barr}{\begin{array}}
\newcommand{\earr}{\end{array}}

\def\gs{\mathrel{
   \rlap{\raise 0.511ex \hbox{$>$}}{\lower 0.511ex \hbox{$\sim$}}}}
\def\ls{\mathrel{
   \rlap{\raise 0.511ex \hbox{$<$}}{\lower 0.511ex \hbox{$\sim$}}}}

\title{Generation of Neutrino mass from new physics at TeV scale and
Multi-lepton Signatures at the LHC}

\author{Gulab Bambhaniya,}
\author{Joydeep Chakrabortty,}
\author{Srubabati Goswami,}
\author{and Partha Konar}
 
\affiliation{Physical Research Laboratory, Ahmedabad-380009, India}
 
\emailAdd{gulab@prl.res.in}
\emailAdd{joydeep@prl.res.in}
\emailAdd{sruba@prl.res.in}
\emailAdd{konar@prl.res.in}


\abstract
{
In this paper we consider generation of naturally small 
neutrino masses from a dimension-7 operator. 
Such a term can arise in presence of a scalar quadruplet 
and a pair of vector-like fermion triplets 
and enables one to obtain small neutrino masses through TeV scale
linear seesaw mechanism. 
We study the phenomenology of the charged scalars of this model,
in particular, the multi-lepton signatures at the Large Hadron Collider. 
Of special importance are the presence of the same-sign-tri-lepton signatures originating from the 
triply-charged scalars. The Standard Model 
background for such processes is small and hence this  
is considered as smoking gun signal 
of new physics. We also looked  for events with 
three, four, five and six-leptons which have negligible contamination from 
the Standard Model. We further point out the spectacular lepton flavour violating four-lepton signal which
can be the hallmark for these type of models. We also compute the added contributions in the rate for the 
Standard Model  Higgs decaying to two photons via the charged scalars in this model.
}

\begin{document}
 
 \maketitle

 \newpage
 
\section{Introduction} \label{sec:intro}

The phenomenal results from the CMS and ATLAS detectors at the Large Hadron 
Collider (LHC) have 
conclusively established the  evidence for the existence of a new boson 
with a mass of around $\sim$ 125 GeV \cite{cms,atlas}.  
It is likely that this new particle is the elusive Higgs boson 
which is responsible for giving mass to the fermions and bosons 
through the mechanism of spontaneous breaking of the electroweak
symmetry. Further data on the properties of this particle would 
determine if it is indeed the Standard Model (SM) Higgs
and/or if there is any signature 
of new physics beyond SM. 

In the SM the sole fermion whose mass is not generated by the 
Higgs mechanism is the neutrino. This is due to the absence of 
right handed neutrinos in the SM. However, results from neutrino oscillation experiments were 
able to establish that neutrinos have tiny non-zero masses and consequently
they mix between flavours. 
The parameters governing neutrino oscillation are mass squared differences 
and mixing angles. 
Present oscillation data have conclusively established the 
existence of at least two massive neutrinos by measuring
two independent mass squared differences as,  
$\Delta m^2_{\rm solar} \sim 10^{-5}$ eV$^2$ and $|\Delta m^2_{\rm atm}| 
\sim 10^{-3}$ eV$^2$. This combined with the cosmological bound
on sum of the light neutrino masses, $\Sigma m_i \ls 0.23 $ eV \cite{cosmobound} suggest
that neutrino masses are much smaller than those of the charged leptons. 
Of course, one can just add a gauge singlet right handed 
neutrino to the SM and generate neutrino masses. Then to achieve 
neutrino masses $\sim \sqrt{\Delta m^2_{\rm atm}} \sim 0.05$ eV one needs minuscule 
Yukawa couplings much smaller than that of their charged fermion counterparts. 
There can be  no other signature of these singlet neutrinos apart from 
neutrino oscillations. Phenomenologically more interesting and theoretically more natural 
is the seesaw mechanism in which additional heavy particles are
added to the theory. This is instrumental in lowering the neutrino masses to sub-eV scale.
Seesaw mechanism predicts neutrinos to be of Majorana 
nature. This insinuates lepton number violation (LNV) which leads  
to neutrinoless double beta ($0\nu\beta \beta$) decay. 
It is also possible to generate a lepton asymmetry within the framework
of seesaw mechanism, through leptogenesis, which can then be converted 
into the  observed baryon asymmetry of the universe \cite{Fukugita:1986hr}. 
The origin of seesaw is the effective dimension-5 operator $ \kappa LLHH $ as suggested by Weinberg in ref.~\cite{weinberg},
where $L$ and $H$ denote the SM lepton, and Higgs fields respectively. 
Such an operator can arise through the tree level exchange of 
singlet fermions in the type-I seesaw 
\cite{typeI1,typeI2,typeI3,typeI4,typeI5}, 
$SU(2)$ triplet scalars in the type-II 
seesaw \cite{typeII1,typeII2} and $SU(2)$ triplet fermions in the type-III seesaw 
\cite{typeIII}. These features can be captured in
the framework of the 
Grand Unified Theory (GUT),  
in which the natural seesaw scale is very high.         
The neutrino mass $m_\nu \sim \kappa v^2 $ arises after spontaneous symmetry breaking 
when the Higgs field acquires a vacuum expectation value ($vev$), $v$.  
This effective coupling, $\kappa$, can be related to the Yukawa coupling of the neutrinos,  
$Y_\nu$, as $\kappa \sim Y_\nu^2/M$.  
Consequently, for  $Y_\nu \sim 1$, $v \sim 100$ GeV, and
$m_\nu \sim 0.01$ eV one requires $M \sim 10^{15}$ GeV which is 
close to the GUT scale. 
However, such a high scale of new physics is beyond the reach of the 
colliders. Recently search for new physics beyond SM has got an unprecedented 
boost because of the LHC. Motivated by this there have been several attempts to probe if 
TeV scale new physics, accessible to LHC, can be responsible for generation of neutrino masses.
For $M \sim 1$ TeV in the dimension-5 operator
one would need $Y_\nu \sim 10^{-6}$ to generate the neutrino 
mass in the right ballpark unless specific 
textures for the Yukawa matrices are assumed \cite{smirnov-kersten,ARC,Pilaftsis}. 
For such low values of coupling one does not expect any signal
at colliders.

There are several ways for lowering the seesaw scale to TeV. 
For a recent review we refer to \cite{muchun}. 
One of the ways to generate neutrino mass via new physics at
TeV scale is through higher dimensional operators 
\cite{higher-dim1,dim7,higher-dim3,higher-dim4,higher-dim5,higher-dim6,higher-dim7}. 
These operators typically contain the factor 
$ \sim \frac{v^2}{M} (\frac{v}{M})^{d-5}$ in the expression of neutrino mass. 
This implies the suppression factor $M^{d-4}$ in the denominator, where $d$ is the dimension of the operator. 
Consequently, the cutoff scale of new physics can be lowered to TeV without making the Yukawa couplings 
minuscule. Such operators can arise at tree as well as loop level and requires extension of the SM field content by 
new fermions and scalars belonging to higher representations 
of $SU(2)$. Since the scale of new physics in these models is at TeV, it 
is conceivable that these new particles can be produced and 
studied at the LHC. In particular, ref.~\cite{dim7} proposed the generation of neutrino mass through dimension-7 operator 
and discussed its signatures. Charged lepton flavour violation in this model has been considered in \cite{higher-dim5}.

In this paper we consider the scenario in which 
neutrino mass is generated through 
dimension-7 operator. 
The model contains a pair of vector-like triplet fermions with hypercharge of 
2 units and a scalar with isospin 3/2. 
The charged scalars in this model can give multi-lepton signatures at the LHC. In particular, we study
the same-sign-tri-leptons as well as other multi-lepton signatures with multiplicity higher than two.  
Some of the multi-lepton signatures have also been mentioned in \cite{dim7}.
We  perform a realistic simulation
using CalcHEP and PYTHIA to estimate signal events and 
incorporate the SM backgrounds using  
ALPGEN. In addition, we consider the possibility  
of observing the flavour violating charged lepton signals in the 
multi-lepton events at the LHC, in the context of this model. 
In particular, we concentrate on the  
 lepton flavour violating four-lepton signals.
Total lepton number is conserved in these kind of events although 
lepton flavour violation (LFV) can occur. 
These processes have very low 
background coming from the SM and hence are considered as harbingers of 
new physics \cite{Feldmann:2011zh,Ghosh:2012ag}. 
For recent studies of multi-lepton search at the LHC, see for instance \cite{cms_multilepton, atlas_multilepton}.
A comparative study of multi-lepton search in different seesaw models has been performed in \cite{del-Aguila}.

As is well known, the $H \rightarrow \gamma \gamma$  process constitutes 
the major discovery channel for the Higgs boson because of low  SM background.
The currently observed rate for this by the ATLAS \cite{atlas-2photon} and 
CMS \cite{cms-2photon} collaborations indicates some departure from the 
SM predictions. The model we consider may provide an extra contribution to the di-photon decay rate 
because of the presence of extra charged scalars. 
We include some discussions on this possibility. 

The plan of the paper is as follows. In section~\ref{sec:model} we present the relevant part of the 
Lagrangian due to the additional particles. 
In  the following section we explain the origin of neutrino mass and the neutral fermion mass matrix.
In section~\ref{sec:pheno} we discuss the  
production and decay modes of the charged scalars as well as the 
phenomenological signatures of new physics. 
In the next section we outline the details of the 
simulation and event selection criteria.  
The SM backgrounds for our signal processes
are also estimated. We illustrate the results for multi-lepton signals choosing some
benchmark points. We also present the lepton flavour violating (LFV) four-lepton signal
that can be obtained in the model under consideration. 
Possible deviations in the Higgs to di-photon decay rate, due to the charged scalars are estimated. 
Finally, we summarize and conclude our discussion.
In the appendix we  present the details of the model
and the Feynman rules which are used in our calculations.

\section{The Model} \label{sec:model}
Prime aim of this model is to generate light neutrinos without 
fine tuning in the Yukawa couplings and at the same time having the 
scale of new physics at TeV. 
This requires introduction of  exotic fermions and scalars,
heavier than the SM fields.
The gauge group of the model under consideration is the same as the SM: 
$G=SU(3)_{c} \otimes SU(2)_{L} \otimes U(1)_{Y}$ with 
an enlarged particle content that includes  
the following fields: an isospin 3/2 scalar
\begin{eqnarray}
\Phi = 
\left( \barr{cccc}
\Phi^{+++}   &   \Phi^{++} & \Phi^{+} & \Phi^{0}  \earr \right)_{Y=3}, 
\end{eqnarray}
and 
a pair of  vector-like fermion triplets 
\begin{eqnarray}
\Sigma_{R,L} = \left( \barr {ccc}  
\Sigma_{R,L}^{++} & \Sigma_{R,L}^{+} &  \Sigma_{R,L}^{0}  \earr \right )_{Y=2}.
\end{eqnarray}
Note that although the above fermionic representations have a non-zero hypercharge,
the chiral anomaly gets canceled as they are vector-like by nature.  
The scalar kinetic and potential terms after involving $\Phi$ read as
\begin{eqnarray}
\label{lphi}
{\cal{L}}_{\mathrm scalar}  = 
(D^{\mu} \Phi )^{\dag }(D_{\mu} \Phi) + 
(D^{\mu} H )^{\dag }(D_{\mu} H) +  V(H,\Phi),
\end{eqnarray}
where
\begin{equation}
D_{\mu} S = \left(\partial_{\mu} -i g \vec{T} {\bf .}  \vec{W}_{\mu} - i g' \frac{Y}{2} B_{\mu}\right) S .
\end{equation}
In the above expression $S$ can be either $H$ or $\Phi$.  The generators $T_a$'s are the 
Pauli matrices for $H$, whereas for $\Phi$ these are the $SU(2)$ generators in the 
isospin 3/2 representation, see appendix~\ref{appendix_scalar}.  
The interactions of the new scalar field $\Phi$ with the gauge bosons originate
from the above term.   
The scalar potential is given as
\begin{eqnarray} \nonumber
V(H,\Phi) &=& {\mu^2_H} H^{\dagger} H + {\mu^2_\Phi} {\Phi^{\dagger}} \Phi + \frac{\lambda_1}{2} (H^{\dagger} H)^2 + \frac{\lambda_2}{2} ({\Phi^{\dagger}} \Phi)^2 \\ 
& & +  \lambda_{3} (H^{\dagger} H) ({\Phi^{\dagger}} \Phi) +  \lambda_{4} (H^{\dagger} \tau_{a} H) 
({\Phi^{\dagger}} T_{a} \Phi) + \{ \lambda_{5} H^3 \Phi^{*} + h.c. \}.
\label{potential}
\end{eqnarray}

The electroweak symmetry is broken spontaneously once the Higgs acquires the vacuum expectation value ($vev$), $v$. As the other scalar
$\Phi$ is also non-singlet under $SU(2)_L\otimes U(1)_Y$, the $vev$ of $\Phi$, i.e. $v_{\Phi}$ can also be responsible for this breaking and
affects the $\rho$ parameter of the SM. 
Thus, $v_{\Phi}$ gets constrained from the $\rho$ parameter 
which gets modified as $\rho \approx (1-6 v_{\Phi}^{2}/v^{2})$.
In order to  satisfy  the $3\sigma$ range of this parameter
$\rho= 1.0004^{-0.0012}_{+0.0009}$ 
\cite{Beringer:1900zz}, $v_\Phi$ 
must be less than 2.01 GeV.
Minimisation of potential in eq.~\ref{potential} leads to the induced $vev$ $v_\Phi=-\lambda_{5} \frac{v^3}{M_{\Phi^0}^2 }$, 
where $v= \sqrt{\frac{-\mu_H^{2}}{\lambda_{1}}}$ is $vev$ of the SM Higgs.  
The mass of the neutral scalar $\Phi^0$ comes out to be
\begin{eqnarray}
M_{\Phi^0}^2  = \mu_\Phi^2 + \lambda_3 v^2 + \frac{3}{4} \lambda_4 v^2.
\label{eq:phi_0}
\end{eqnarray}
The mass of $i^{th}$ component of the quadruplet field $\Phi$ with absolute value of electric charge $q_i$ can be expressed 
as{\footnote{Note that we get the splitting parameter ($\Delta{M^2}$) as $\frac{\lambda_4}{2}v^2$ instead of 
$\frac{\lambda_4}{4}v^2$ in reference \cite{dim7}. 
}} 
\begin{eqnarray}
M_{\Phi^i}^2 &=& M_{\Phi^0}^2 - q_i \frac{\lambda_4}{2} v^2 = M_{\Phi^0}^2  - q_i {\Delta{M^2}}.
\label{eq:phi_i}
\end{eqnarray}
${\Delta{M^2}}$  denotes the difference of 
squared masses between any two successive components in the scalar 
quadruplet.

Note that the mass degeneracy between the members of heavy scalar are 
lifted by the $\lambda_4$ coupling once the symmetry is broken.
Choice of this free coupling within the perturbative limit can produce 
successive scalar states from 
a near degenerate to a mass difference of as large as few tens of GeV. 
Depending on the {\it sign of} $\lambda_{4}$, we get two hierarchies in masses of 
the $\Phi$ field:
\begin{eqnarray} \nonumber
M_{\Phi^{0}}>M_{\Phi^{\pm}}>M_{\Phi^{\pm \pm}}>M_{\Phi^{\pm \pm \pm}} \mbox{~for ~positive ~sign ~of} ~{\lambda}_{4}~\mbox{and}~ \Delta{M} -ve, \\
M_{\Phi^{\pm \pm \pm}}>M_{\Phi^{\pm \pm}}>M_{\Phi^{\pm}}>M_{\Phi^{0}} \mbox{~for ~negative ~sign ~of} ~{\lambda}_{4}~\mbox{and}~ \Delta{M} +ve,
\end{eqnarray}
where the mass difference between two successive members is 
constrained as $1.4 ~\mbox{GeV} < |\Delta{M}| < 38 ~\mbox{GeV}$. The lower bound comes from precision electroweak corrections
\cite{Cirelli:2005uq} and the upper bound is for compliance with $\rho$ parameter \cite{Einhorn:1981cy}.
The physical scalar spectrum in this model consists of one triply-charged 
scalar, one doubly and one singly charged scalars as well as three
neutral scalars (two CP even and one CP odd).

\section{Generation of Neutrino Masses} 
\label{sec:neutrino_masses}

In this section we discuss neutrino mass generation 
and the source of lepton number violation. 

\subsection{Neutral Fermion Mass Matrix} \label{sec:mass_matrix}

In this model the neutrino mass ($m_\nu$) comes from the renormalisable Lagrangian\\
\begin{equation} 
{\cal L}_{m_\nu} 
= Y_i  \overline{{l_{iL}}^{C}} H^*\Sigma_L + 
{{Y_i^\prime}} \overline{\Sigma_R} \Phi l_{iL} 
+ \overline{\Sigma_R} M_{\Sigma }  \Sigma_L  
+ h.c., 
\label{nuL} 
\end{equation}
where $Y_i,~{Y_i^\prime}  $ are Yukawa couplings and $i$ is generation index.
The detailed structure of the Yukawa interactions are given in appendix~\ref{appendix_feynman_rules_Yukawa}.
From the Lagrangian (\ref{nuL}) the neutral lepton mass matrix can be 
written in the $(\nu_L, \Sigma^{0}_L, (\Sigma^{0}_R)^C)$ basis as 
\begin{equation}
M_0 =
\begin{array}{ccc}
\begin{pmatrix}
0 & m &  {m^\prime}^{T} \\
m^T & 0 & M_\Sigma \\
{m^\prime} & M_\Sigma & 0 \\
\end{pmatrix}.
\end{array}
\label{neutralmass} 
\end{equation}
In terms of the Yukawa couplings, $m= - Y v$ and ${m^{\prime T}} = {Y^\prime} v_\Phi$.
Since we have introduced a vector-like fermion pair the above mass 
matrix is $5 \times 5$ and is of rank 4. Thus, the neutral fermion spectrum 
consists of 
two nearly degenerate 
heavy neutrinos, one massless neutrino as well as two massive light neutrinos. 

This model contains $SU(2)$ triplet fermions 
as in the type-III seesaw model. However, here they  
have a hypercharge $Y=2$ and are vector-like whereas the
type-III seesaw model 
contains triplet fermion with $Y=0$.  
These two models are different 
in the way neutrino masses are being generated and lepton 
number is being violated. 
In general, the presence of light Majorana neutrinos 
demands lepton number violation (LNV) by two units,
i.e. $\Delta L=2$.  In conventional 
type-III seesaw mechanism, the triplet fermions which are being integrated out 
during the seesaw process 
have Majorana mass term and the violation of 
lepton number  is directly reflected from the mass insertion in 
the propagator. But in this case the triplet fermions are vector-like. Thus,
they have Dirac type mass term rather than the Majorana mass. 
However, lepton number violation can come from the $\bar{\Sigma} \Phi l$  
term in the Lagrangian if we assign a lepton number of +1 to the $l$ field 
and -1 to the vector-like $\Sigma_L$ and $\Sigma_R$ fields\footnote{It is also possible to assign lepton number +1 to the $\Sigma_L$ and
$\Sigma_R$ fields, in which case lepton number will be violated at the 
$\overline{l^C} H^* \Sigma$ vertex.}.

\begin{table}[htb]\centering
\begin{tabular}{|c|c|c|c|c|}
\hline      
$ v_{\Phi}$ (GeV) & $y$ & $y'$ & $m_{\nu}$ (eV) \\
\hline
$5\times10^{-6}$  & $10^{-1}$  & $10^{-3}$ & 0.05  \\
\hline
$5.1\times10^{-5}$  & $10^{-1}$  & $10^{-4}$ & 0.05  \\
\hline
$4.3\times10^{-5}$ & $10^{-1}$  & $10^{-4}$ & 0.04  \\
\hline
$0.5 $ &  $ 10^{-3} $ & $10^{-6}$ & 0.04  \\
\hline
\end{tabular} 
\caption{\label{numassorder} Order of neutrino mass for  $v = 174$ GeV, $M_{\Sigma}= 3500$ GeV, $M_{\Phi^0}= 400$ GeV.}
\end{table}

Note that with the above assignment of lepton number the matrix in eq.~\ref{neutralmass} has 
the form of the linear seesaw mass matrix \cite{linear01, linear1, linear2}.
A naturally small neutrino mass can be generated in this model assuming 
small lepton number violation. 
Expressing the Yukawa coupling matrices  $Y = y \hat{\bf{y}}$ and 
$Y^\prime = y^\prime \hat{\bf{y^\prime}}$ in terms of the unit vectors 
$\hat{\bf{y}}$ and $\hat{\bf{y^\prime}}$,   
the mass matrix can be fully reconstructed in terms of 
the oscillation parameters apart from the overall coupling constants
$y$ and $y^\prime$ \cite{gavelamfv}. 
These coupling strengths 
can be constrained from the consideration of vacuum stability bound of the 
electroweak vacuum, LFV in charged lepton decays 
and neutrino oscillation results \cite{khan}.

The above matrix can be diagonalized in the limit $M_\Sigma >> m, {m^\prime}$ 
to generate the light neutrino mass matrix 
$m_{\nu}$
using the seesaw approximation. To the leading order this can be 
expressed as, 
\begin{eqnarray}
m_{\nu} = - m \frac{1}{M_\Sigma} m^\prime  -  {m^\prime}^T \frac{1}{M_\Sigma} m^T. 
\label{eq:mnu} 
\end{eqnarray}
Then the expression for $m_{\nu}$ can be written as, 
\beqa 
(m_{\nu})_{ij}  =  
\frac{(Y_i Y_j^{\prime} +  {Y_i^\prime} Y_j) v_\Phi v}{M_\Sigma} =
-\frac{\lambda_5(Y_i Y_j^\prime +  {Y_i^\prime} Y_j) v^4}{(M_{\Sigma}
M_{\Phi^0}^2 )}.
\label{eq:mnu2} 
\eeqa
Note that the  neutrino mass goes to zero in the limit the lepton 
number violating coupling $Y^\prime$ goes to zero. Thus, 
a naturally small neutrino mass can be generated assuming 
small lepton number violation.
Along with the lepton number violation it is also possible to have lepton flavour violation 
in this model. This is reflected in the effective vertex ($\Phi \ell \ell$), 
as shown in appendix~\ref{appendix_feynman_rules_effective_vertex}.

In our analysis we consider $v_{\Phi}$ and $M_{\Phi^0}$ to be independent parameters with $\lambda_5<0$.
In table~\ref{numassorder} we present the typical values of 
$v_\Phi$ used in our analysis and the corresponding values of 
$y$ and $y^\prime$ in order to generate correct order of magnitude for 
the neutrino mass  for the representative values of $M_{\Phi^0}$ 
and $M_\Sigma$.

\subsection{Origin of Neutrino Mass} 
\begin{figure}[htb]
\begin{center}
\includegraphics[width=7cm]{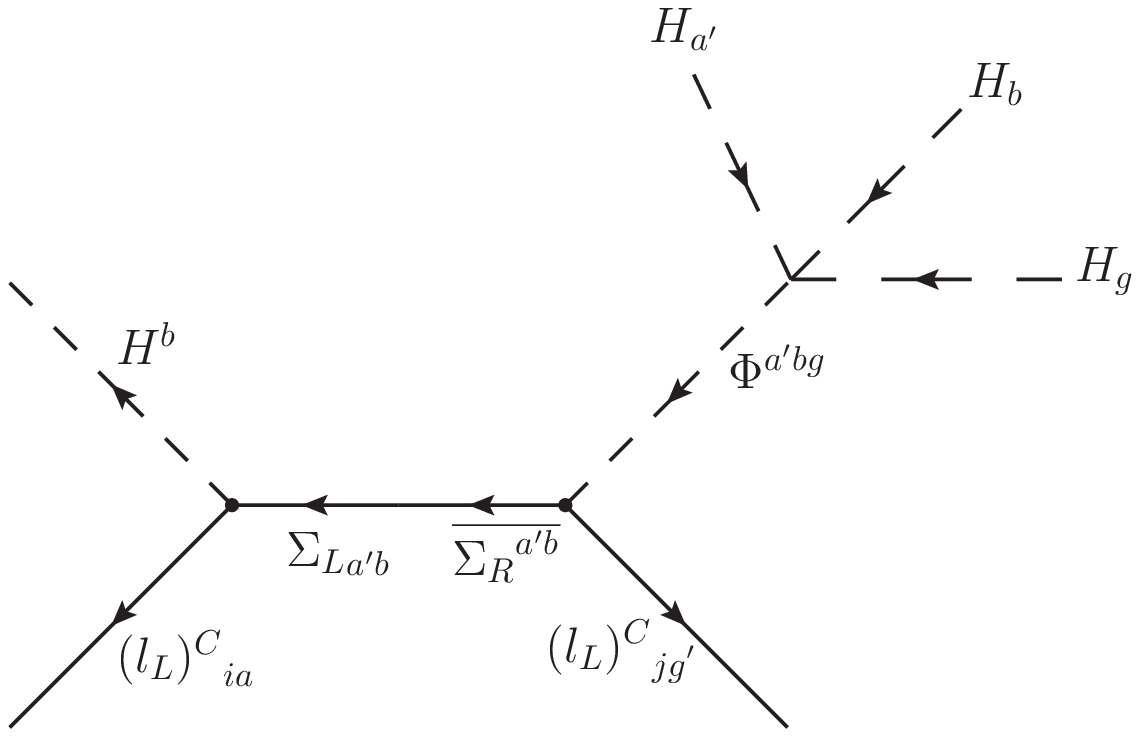}
\includegraphics[width=7cm]{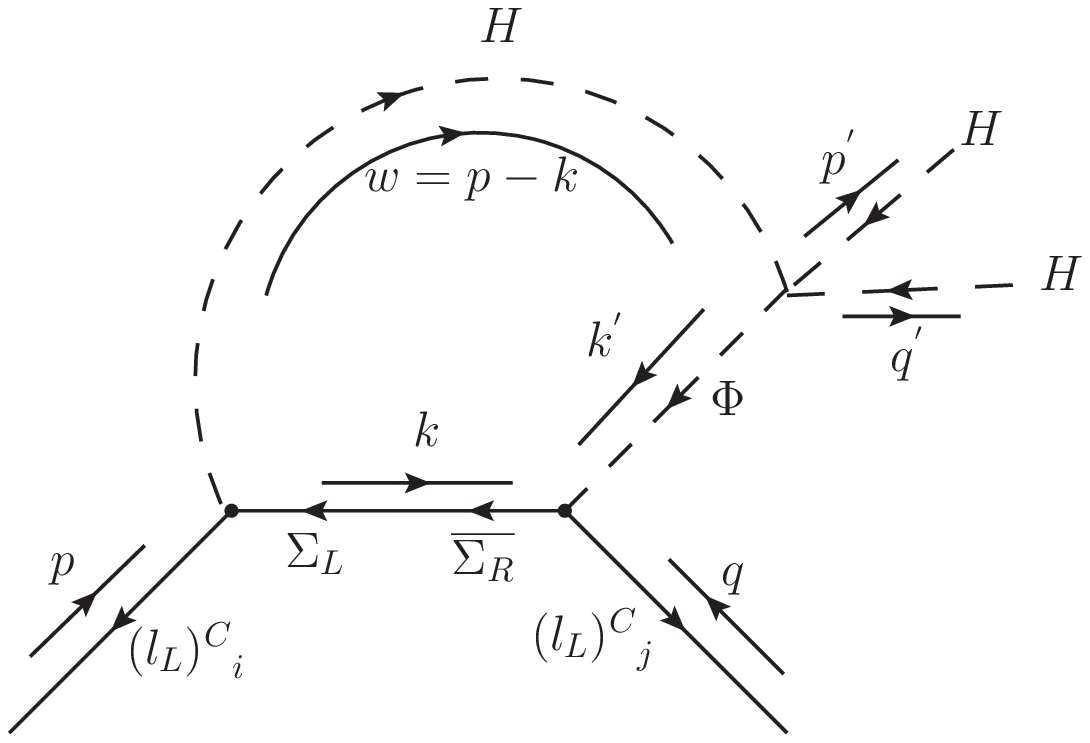}
 \caption{Tree level diagram (left panel) generating dimension-7 seesaw
 operator  and 1-loop diagram (right panel) generating dimension-5 operator 
for neutrino masses. }
 \label{fig:loop}
 \end{center}
 \end{figure}

The tree level diagram from which the neutrino mass
originates is given in figure~\ref{fig:loop}(left panel).
After integrating out heavy fermion fields $\Sigma$,
$\overline{\Sigma}$ and scalar field $\Phi$, this diagram gives rise to a 
dimension-7 effective Lagrangian
\beqa
{\cal L}_\kappa
&=& - \kappa_{ij} \left( {\overline{l_L^C}}^i_a
H_{a^{'}}  l_{Lg^{'}}^j H^{b}H_{b}H_{g} \right) \varepsilon_{aa^{'}} \varepsilon_{gg^{'}} + {\rm h.c.} \; \; ,
\label{eq:L-kappa-1}
\eeqa
where 
\begin{equation}
\kappa_{ij} = -\frac{(Y_i Y_j^{'} + Y_i^{'} Y_j) \lambda_5 } {M_{\Sigma}{M_{\Phi^0}^2}},
\label{eq:kappa} 
\end{equation}
which after spontaneous symmetry breaking generates the  
neutrino mass  given in
eq.~\ref{eq:mnu2}. The details of the  calculation for obtaining 
eq.~\ref{eq:kappa} are given in the appendix~\ref{appendix_feynman_rules_effective_mass}.

In order to get dominant contribution for neutrino mass from the dimension-7 operator 
one needs to forbid the dimension-5 terms. This is ensured by the absence of singlet fermions,
$Y=0$ triplet fermions and triplet scalars in the model.
However, dimension-5 operator can arise at the 1-loop level through diagram
depicted in the right panel of figure \ref{fig:loop}.

\begin{figure}[htb]
\begin{center}
\includegraphics[width=8cm, height=7cm]{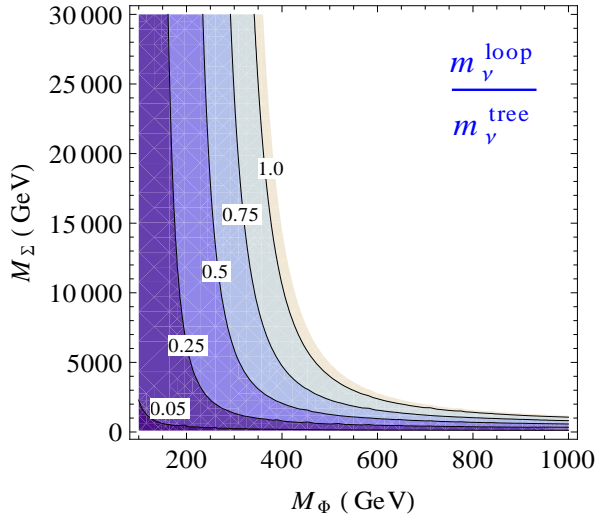}
 \caption { Contour plot of the ratio $m_{\nu}^{loop}/m_{\nu}^{tree}$ 
in the $(M_\Phi - M_\Sigma)$ plane. } 
 \label{fig:contour}
 \end{center}
 \end{figure}

 Including the above diagram the total contribution to $m_\nu$ becomes 
$m_\nu^{total} = m_{\nu}^{tree} + m_{\nu}^{loop} $,
where $m_{\nu}^{tree}$ is given by eq.~\ref{eq:mnu2}. 
The loop contribution to the neutrino mass, $m_{\nu}^{loop}$, can be computed as
\begin{equation} 
\label{m_loop}
(m_{\nu})_{ij}^{loop}=\frac{\left(3+\sqrt{3}\right) \lambda _5 v^2 M_{\Sigma } 
\left(Y_i Y_j^{'}+ Y_i^{'} Y_j \right)}{16 \pi ^2 \left(M_{\Phi }^2-M_H^2\right)}
\left(
\frac{M_{\Phi}^2 \log \left(\frac{M_{\Sigma }^2}{M_{\Phi }^2}\right)} {M_{\Sigma }^2-M_{\Phi }^2}-
\frac{M_H^2 \log \left( \frac{M_{\Sigma }^2}{M_H^2}\right)} {M_{\Sigma}^2-M_H^2}
\right),
\end{equation}
where $M_H$ is mass of the SM Higgs.

Note that only $\Phi^0$ and $\Phi^+$ will contribute to the loop diagrams in figure~\ref{fig:loop}. 
In deriving eq.~\ref{m_loop} we have assumed these two states are degenerate with mass $M_\Phi$.

In figure~\ref{fig:contour} we make a contour plot of the ratio 
$m_{\nu}^{ loop}/m_{\nu}^{ tree}$ 
in the $(M_\Phi - M_\Sigma)$ plane. 
From the plot it is clear that for smaller values of 
$M_{\Phi}$ and $M_\Sigma$, 
the dimension-7 contribution dominates over that 
coming from the dimension-5 term. 
This is the region relevant for our study and thus it 
suffices to take only the tree level contribution.

\subsection{Neutrino Mixing Matrix} 
The light neutrino mass matrix $m_\nu$ can be diagonalized in the basis where the charged lepton mass matrix 
is diagonal as:
\begin{eqnarray}
\label{diagonal}
U_{PMNS}^T m_\nu U_{PMNS} = m_\nu^{diag},
\end{eqnarray}
where $m_\nu^{diag} 
=\mbox{diag}(m_1,m_2,m_3)$, and $U_{PMNS}$ is the neutrino mixing matrix. 

Since in this model the smallest mass is zero, one can express the 
two other mass eigenvalues in terms of the mass squared differences
($\Delta m^2_{ji} \equiv m_j^2 - m_i^2$) 
governing solar and atmospheric neutrinos as:
\begin{itemize} 
\item Normal Hierarchy (NH) : $m_1 << m_2 \approx m_3$  
\begin{equation} 
m_1 = 0,\quad
m_2 = \sqrt{\Delta m^2_{21}}  \quad, 
m_3 = \sqrt{\Delta m^2_{32}+\Delta m^2_{21}},   
\end{equation}

\item Inverted Hierarchy (IH) : $ m_3 << m_1 \approx m_2$  
\begin{equation} 
m_3 =0, \quad 
m_1 = \sqrt{\Delta m^2_{13}}  \quad, 
m_2 = \sqrt{\Delta m^2_{21} +\Delta m^2_{13}}.
\end{equation}
\end{itemize}

\begin{table}[htb]\centering
  \begin{tabular}{|c|c|c|c|}
    \hline
    parameter & best-fit & 3$\sigma$ range & Values used by us
    \\
    \hline
    $\Delta m^2_{21}\: [10^{-5}~\text{eV}^2]$
    & 7.62 & 7.12--8.20 & 7.59\\[1.5mm] 
    $\sin^2\theta_{12}$
    & 0.320 & 0.27--0.37& 0.33\\[2.5mm]  
    $|\Delta m^2_{31}|\: [10^{-3}~\text{eV}^2]$
    &
    \begin{tabular}{c}
      2.55 (NH) \\
      2.43 (IH)
    \end{tabular}
    &
    \begin{tabular}{c}
      $2.31-2.74$ (NH) \\
      $2.21-2.64$ (IH) 
    \end{tabular}
   & 2.50
    \\[4.5mm]
    $\sin^2\theta_{23}$
    &
    \begin{tabular}{c}
      0.613 (NH) \\
      0.600 (IH) 
    \end{tabular}
&
    \begin{tabular}{c}
      0.36--0.68 (NH) \\
      0.37--0.67 (IH)
    \end{tabular}
& 0.5
    \\[4mm]
    $\sin^2\theta_{13}$
    &
    \begin{tabular}{c}
      0.0246 (NH) \\
      0.0250 (IH)
    \end{tabular}
    &
    0.017--0.033 & 0.0251\\[4mm]
    $\delta$
   &
   \begin{tabular}{c}
 $0.80\pi$ (NH) \\
     $-0.03\pi$ (IH) 
   \end{tabular}
   &
   $0-2\pi$ & 0\\
       \hline
     \end{tabular}
\caption{ \label{table-osc} The best fit values and $3\sigma$ ranges of neutrino
       oscillation parameters from global analysis of current data 
\cite{valleosc}. The last column contains the oscillation parameters that we have used in our analysis.}
\end{table}

$U_{PMNS}$ in this case is parametrised by three mixing angles 
$\theta_{12}$, $\theta_{23}$ and $\theta_{13}$,
one Dirac phase ($\delta$) and one Majorana phase ($\alpha$) as:
\begin{equation}
U   =  \left(
 \begin{array}{ccc}
 c_{12} \, c_{13} & s_{12}\, c_{13} & s_{13}\, e^{-i \delta}\\
 -c_{23}\, s_{12}-s_{23}\, s_{13}\, c_{12}\, e^{i \delta} &
 c_{23}\, c_{12}-s_{23}\, s_{13}\, s_{12}\,
e^{i \delta} & s_{23}\, c_{13}\\
 s_{23}\, s_{12}-\, c_{23}\, s_{13}\, c_{12}\, e^{i \delta} &
 -s_{23}\, c_{12}-c_{23}\, s_{13}\, s_{12}\,
e^{i \delta} & c_{23}\, c_{13}
 \end{array}
 \right) P \, ,
\label{upmns_param}
\end{equation}
with $c_{ij} = \cos\theta_{ij}$ and $s_{ij} = \sin\theta_{ij}$ and $P=\mbox{diag}\{1,1,e^{i\alpha}\}$.
The oscillation parameters (best-fit values  and  $3\sigma$  ranges) and the 
illustrative values chosen by us, are summarized in table~\ref{table-osc}.
In our analysis we have taken the phases $\delta$ and $\alpha$ to be zero.


\section{Phenomenological Implications} \label{sec:pheno}

This model provides an avenue to test the mechanism of neutrino mass 
generation at the LHC. 
The presence of the isospin 3/2 scalar multiplet $\Phi$, 
specially the triply- and doubly-charged scalars can give rise to rich 
phenomenology at the LHC. 
Specifically, the cascade decays of these heavy charged scalars
lead to multi-lepton final states. 
In this section we discuss in  detail the production and decay modes of 
these scalar fields and possible signals at the LHC.

\subsection{Production and Decay of Isospin 3/2 Scalar}

\begin{figure}[htb]
\includegraphics[width=6.5cm,angle=0]{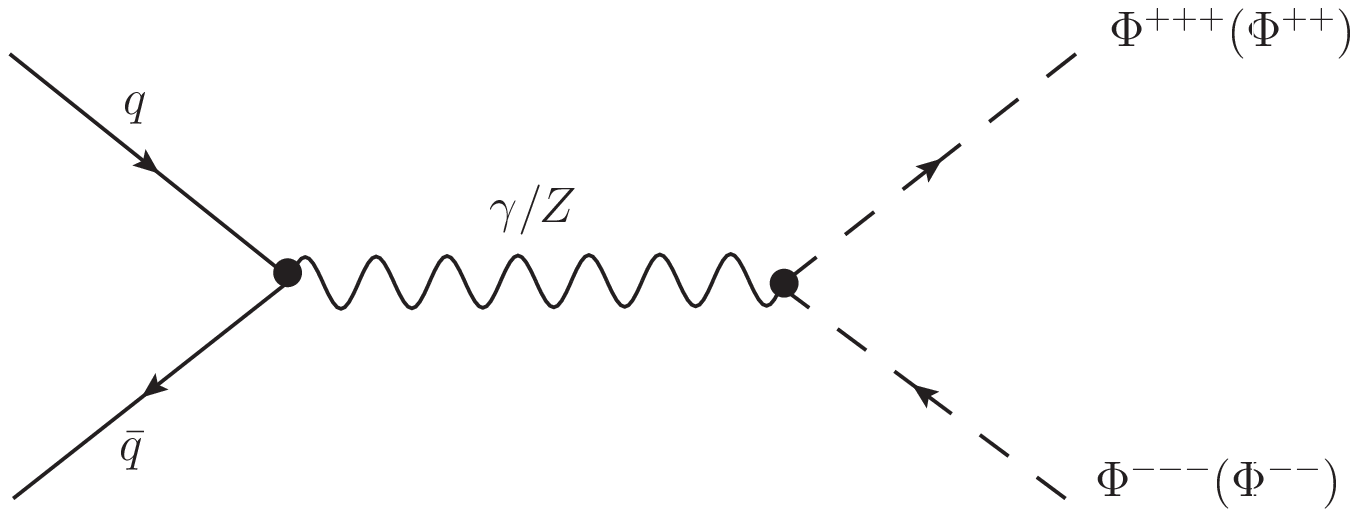}
\hskip 2cm 
\includegraphics[width=6.5cm,angle=0]{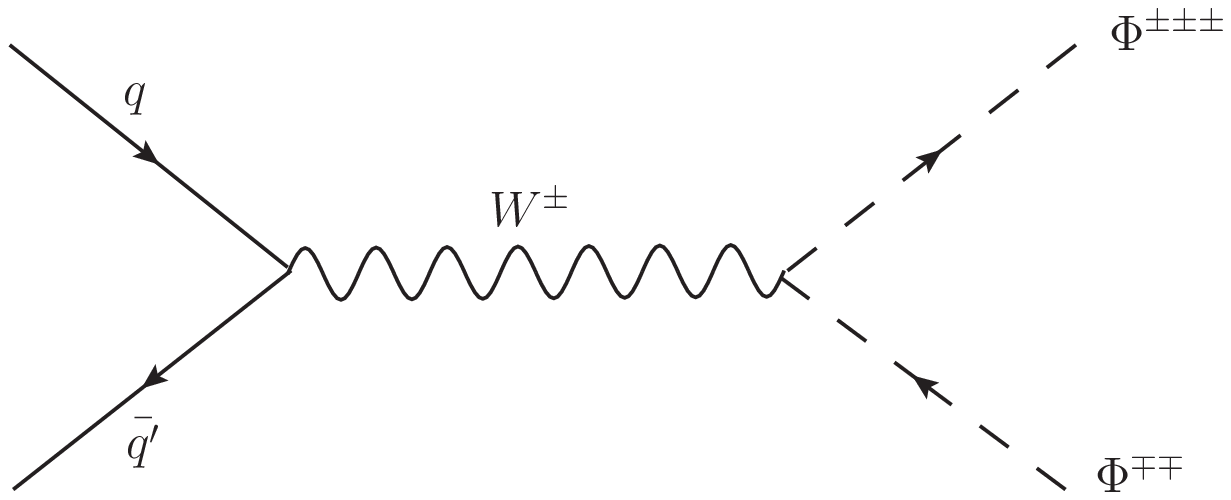}
\includegraphics[width=6.5cm,angle=0]{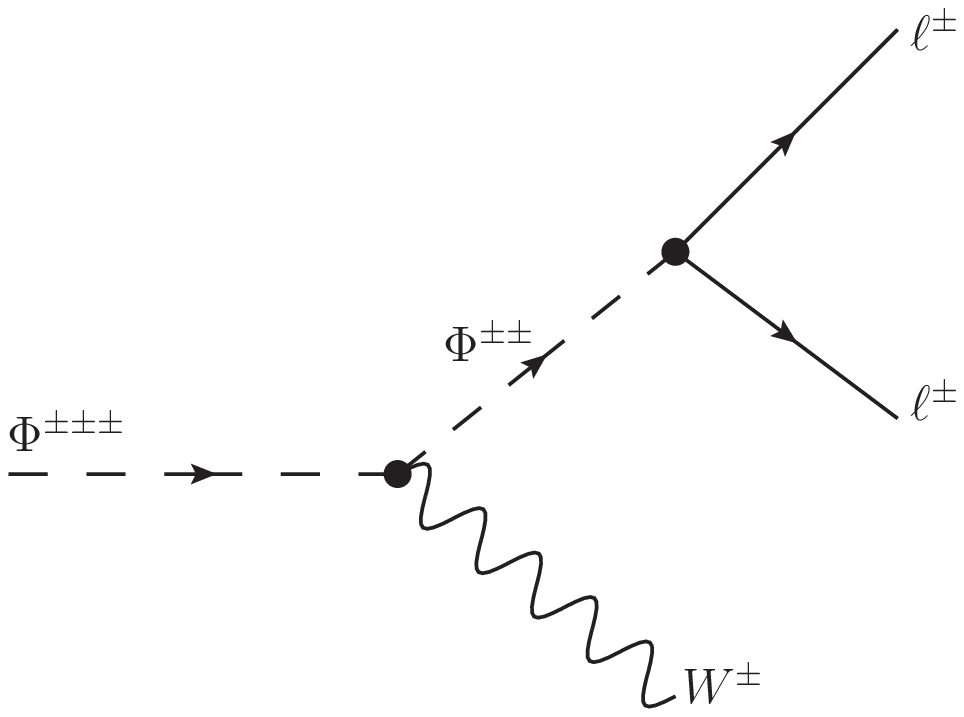}
\hskip 2cm
\includegraphics[width=6.5cm,angle=0]{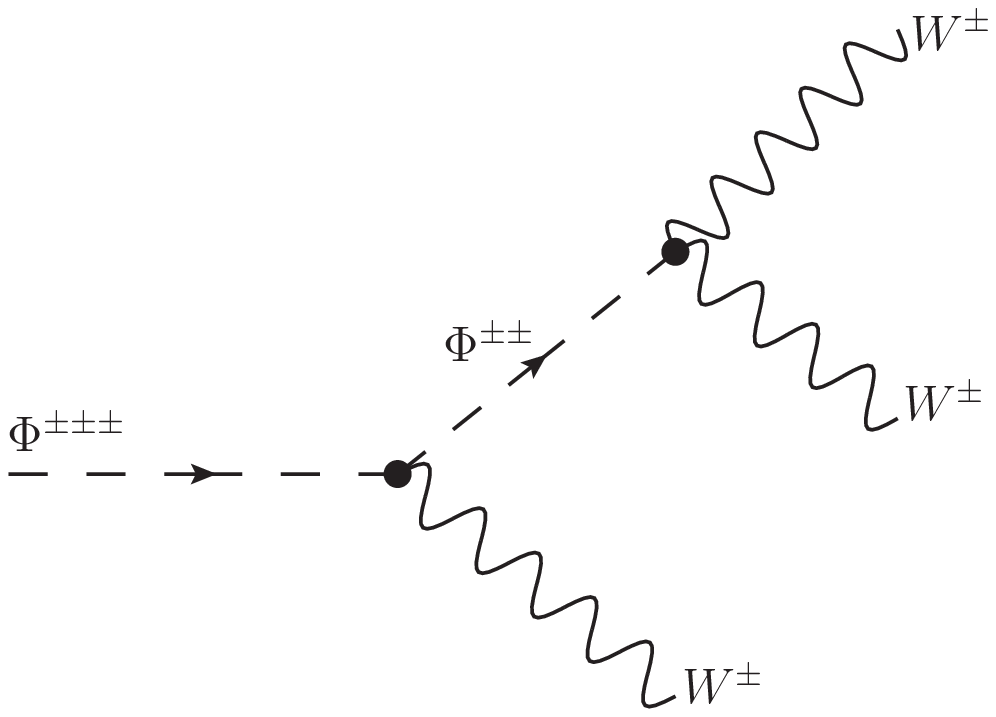}

\caption{\small Representative diagrams for production and decay of doubly- and triply-charged 
scalars at hadron collider leading to multi-lepton ($\ge 3$) final states.}
\label{fig:production_decay}
\end{figure}

The heavy scalars are produced in pair through electroweak gauge bosons at Large Hadron Collider through the following processes\footnote{
Note that $\Phi^{0}\Phi^{0}$ production is absent due to the lack of coupling between gauge bosons and pair of neutral scalars.}:
\begin{eqnarray}
pp & \xrightarrow{Z/\gamma} & \Phi^{\pm\pm\pm}\Phi^{\mp \mp \mp},\; \Phi^{\pm\pm}\Phi^{\mp \mp},\; \Phi^{\pm}\Phi^{\mp}; \nonumber \\
pp & \xrightarrow{W^{\pm}} & \Phi^{\pm\pm\pm}\Phi^{\mp \mp},\; \Phi^{\pm\pm} \Phi^{\mp},\; \Phi^{\pm} \Phi^{0}.
\label{produxn} 
\end{eqnarray}
The parton level (lowest-order) representative diagrams contributing to these processes at the LHC are shown in the upper row of 
figure~\ref{fig:production_decay}.

\begin{figure}[htb]
\begin{center}
\includegraphics[width=5.cm, angle=-90]{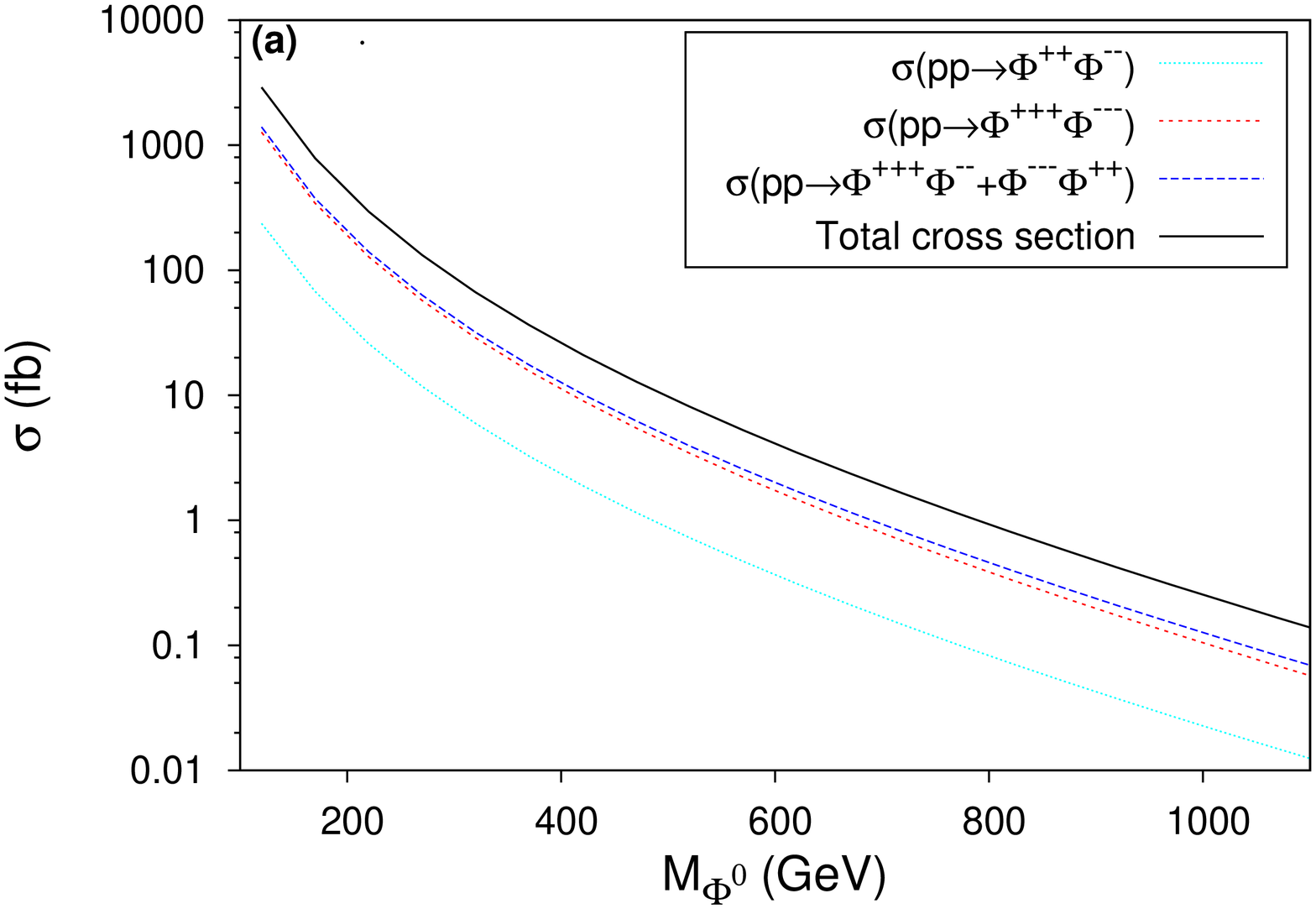}
\; \; \; 
\includegraphics[width=5.cm,angle=-90]{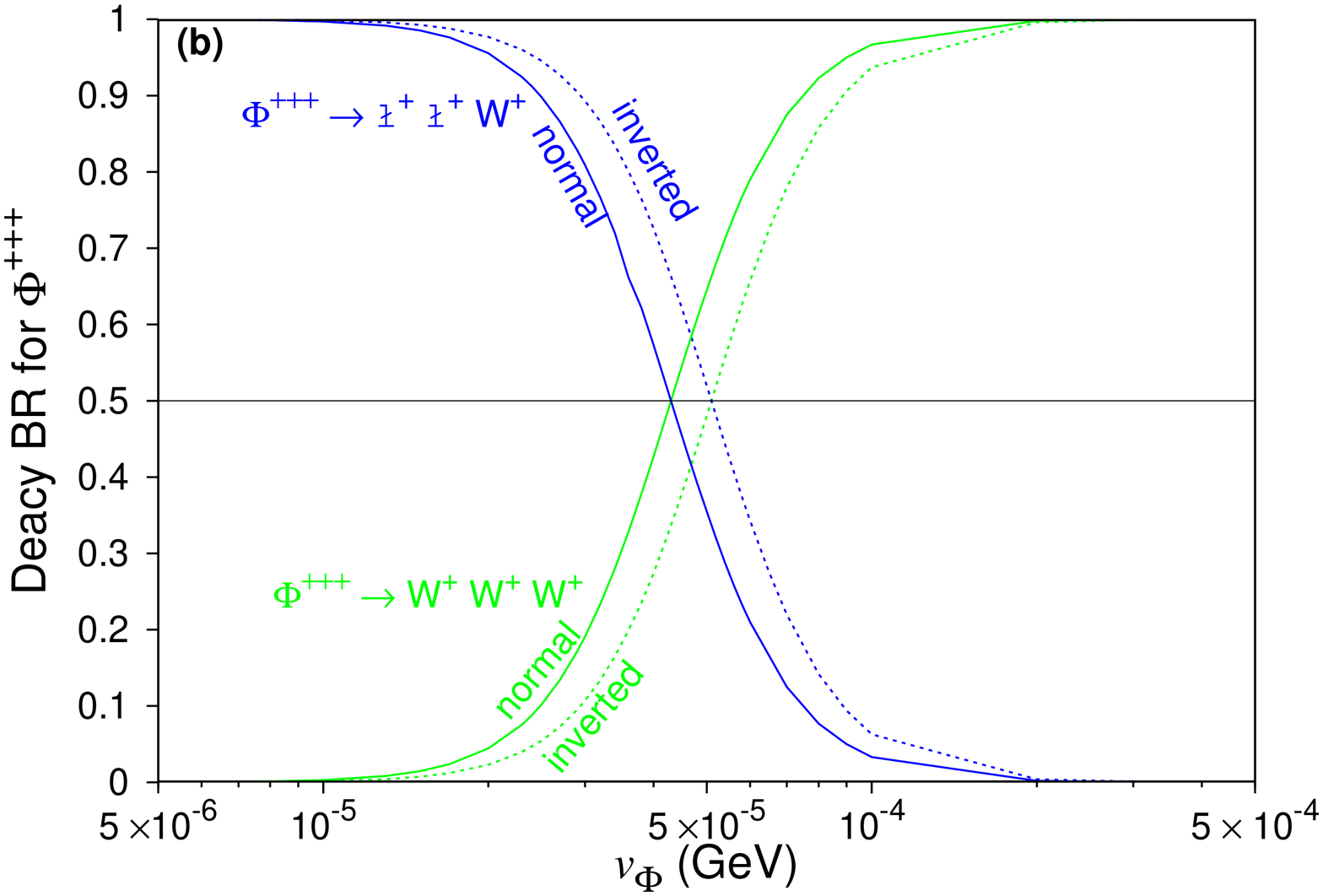}
\caption{Figure (a): Production cross-sections for $pp \xrightarrow{} \Phi^{+++}\Phi^{---}, \Phi^{++}\Phi^{--}, 
\Phi^{\pm \pm \pm}\Phi^{\mp \mp}$ at the LHC with $\sqrt{s}$=14 TeV and $\Delta M= -2.8$ GeV.
Figure (b): Dependence of decay Branching Ratio (BR) of $\Phi^{\pm\pm\pm}$ 
on $v_\Phi$ for IH and NH. Here $\l$ implies all three charged leptons ($e,\mu,\tau$).}
\label{fig:Xsection_h3_decay}
\end{center}
\end{figure}

Figure~\ref{fig:Xsection_h3_decay}(a) shows the production cross-section of the charged scalars at the LHC for 
center of mass energy $\sqrt{s}$ = 14 TeV as a function of the scalar mass parameter $M_{\Phi^0}$. 
We consider the triply- and doubly-charged scalars
which are  expected to give dominant contribution for the multi-lepton signals that we have studied.
We have calculated our hard-scattering matrix elements for parton
level processes by implementing the model in  CalcHEP (version 3.2) \cite{calchep}. 

The doubly charged scalar $\Phi^{\pm \pm}$ can dominantly decay into two $W$-bosons of same charge. 
However, another dominant and in fact remarkable decay channel can be realised from dimension-7 seesaw 
operator generating the neutrino mass. This effective vertex, discussed in 
appendix~\ref{appendix_feynman_rules_effective_vertex} is proportional to 
neutrino mass matrix elements ($m_{\nu_{ij}}$) and $\Phi^{\pm \pm}$ couples to lepton pair ($\ell_i \ell_j$) of same charge leading to 
lepton number violation. Since this vertex depends on the neutrino mass matrix elements, 
one expects relative differences in the signals for normal and inverted neutrino mass hierarchies.

Interplay of these two decay processes controls the
significance of the observed lepton signal as we will demonstrate later.
The triply-charged scalars $\Phi^{\pm \pm \pm}$ can decay into doubly charged scalars 
$\Phi^{\pm \pm}$ associated with $W$-boson apart from other 3-body modes which are suppressed. 
However, narrow mass difference between charged scalars as discussed in section~\ref{sec:model}, 
typically produces off-shell $\Phi^{\pm \pm}$ which can decay further\footnote{
At this point we note that a significant number of $\Phi^{\pm\pm}$ are produced off-shell and thus 
$M_{\Phi^{\pm \pm}}$ cannot be reconstructed from the same-sign-di-lepton invariant mass.}.
The lower row of figure~\ref{fig:production_decay} demonstrates these decay modes of the triply-charged 
scalars. The final decay products ($\l{} \l{} W$ or $WWW$) are determined by the 
corresponding decay channels of the $\Phi^{\pm \pm}$. 
All necessary Feynman rules used in these calculations  are listed in appendix~\ref{feynmanrules}. 

Figure~\ref{fig:Xsection_h3_decay}(b) demonstrates the decay Branching Ratios (BRs) of the 
triply-charged scalars in different modes for both neutrino mass hierarchies NH and IH. This plot is generated
considering neutral scalar mass $M_{\Phi^0} = 400$ GeV 
together with mass difference between two successive scalars $\Delta M = (M_{\Phi^{\pm \pm \pm}} - M_{\Phi^{\pm \pm}})$ = $-2.8$ GeV\footnote{
Since members of $\Phi$ are allowed to have a small mass splitting, mass hierarchies among them, depending upon $sign$ of $\lambda_4$,
would have little impact on production and phenomenological signatures. So our choice of $\Delta M$ is representative by nature.}.
This figure reflects how the interplay of the two decay channels of $\Phi^{\pm\pm}$, for different choices of $v_\Phi$,
affects the BRs of $\Phi^{\pm\pm \pm}$.
Note that the BRs of $\Phi^{\pm\pm\pm}$ closely 
follow that of $\Phi^{\pm\pm}$ excepting for an {\it offsetting factor due to off-shell phase-space} production in the decay of the former.  
From this figure three clear limits emerge: 
\begin{itemize}
\item For small $v_\Phi$ ($\ls 10^{-6}$ GeV), $\Phi^{\pm \pm \pm}$ mostly decays into 
$\l{} \l{} W$ with BR nearly equal to one for both NH and IH.

\item On the other hand, in the larger $v_\Phi$ region ($\gs 10^{-4}$ GeV), $\Phi^{\pm \pm \pm}$ mostly decays into
$3W$ since large values of $v_\Phi$ suppress the lepton number violating effective coupling.

\item For intermediate values of $v_\Phi$, both channels can govern the decay. 
However, the exact values of the branching ratios depend on the neutrino mass hierarchies.
\end{itemize}

The above observations dictate the choice of the benchmark points 
in our study which are listed in table~\ref{choice}.
It can be noted that for lower values of $M_\Phi$ and intermediate ranges of $v_\Phi \sim \mathcal{O}(10^{-5}-10^{-4})$
total decay width of charged scalar can be sufficiently low to show displaced vertex at the detector. 
This can lead to non-pointing multi-lepton signals, although we are not considering such scenarios in our analysis.

\begin{table}[htb]\centering
\begin{tabular}{|c|c|c|}
\hline
Neutrino   & Dominant $\Phi^{\pm \pm \pm}$  &  Choice of \\
hierarchy  & decay  modes                   & $v_{\Phi}$ (GeV)  \\
\hline
IH 
& 
\begin{tabular}{c}
$\l \l W $\\
$\l \l W - W W W$\\
$ W W W$
\end{tabular}
&
\begin{tabular}{c}
$5 \times 10^{-6} $\\
$5.1 \times 10^{-5}$\\
$0.5$
\end{tabular}\\
\hline
NH 
&
\begin{tabular}{c}
$\l \l W$\\
$\l \l W - W W W$\\
$W W W$
\end{tabular}
&
\begin{tabular}{c}
$5 \times 10^{-6} $  \\
$4.3 \times 10^{-5}$  \\
$0.5$
\end{tabular}\\
\hline
\end{tabular}
\caption{Choice of parameter $v_{\Phi}$ for $M_{\Phi^0}= 400$ GeV and $\Delta{M}  = -2.8$ GeV. 
Here $\l$ denotes all three charged leptons ($e,\mu,\tau$).}
\label{choice}
\end{table}

\subsection{Signatures of New Physics} 

The pair productions and subsequent decays of the charged scalars followed by $W^{\pm}$ decay 
lead to different leptonic final states. 
We have considered the following signatures of new physics:
\begin{itemize} 
\item Multi-lepton events -- $~3\ell$, $~ 4\ell$, $~5\ell$ and $~ 6\ell$ events,
\item Same-sign-tri-lepton events $(\mbox{SS}3\ell)$,
\item Lepton flavour violating (LFV) 4 lepton events. 
\end{itemize} 
Here $\ell$ implies only first two generations of charged leptons (e, $\mu$). 
Specially important in this respect is the 
same-sign-tri-lepton signature  
which has very small background coming from SM. 

This model also accommodates spectacular lepton flavour violating decays of the charged scalars. 
These LFV signals can be originated from on-shell and/or off-shell leptonic
decays of $\Phi^{\pm \pm}$. Of particular importance are the LFV four-lepton signal. 
In our study we consider two kind of signals $\ell_i^+ \ell_i^+ \ell_j^- \ell_j^-$ 
or $\ell_i^{\pm} \ell_j^{\pm} \ell_j^{\mp} \ell_j^{\mp}$ ($\ell_i \neq \ell_j=e,\mu$). 
The first one depends on diagonal terms of the light neutrino mass matrix, 
$m_{\nu}$. The other final state is governed by both diagonal and off-diagonal elements of $m_{\nu}$. These LFV signals are not
accompanied by any missing neutrino, and therefore, are expected to be free from SM background.

\section{Results and Simulation} 
\label{sec:results}

\subsection{Simulation and Event Selection} 
\label{sec:simulation-events}
First we implement the model under scrutiny in CalcHEP\cite{calchep}. 
Our signal processes, constructed out of parton level calculation of hard-scattering matrix elements, and 
relevant decay branching ratios are computed for benchmark parameters discussed earlier. 
To perform the full analysis, the ``Les Houches Accord'' (LHA) event file \cite{LHA} 
generated through CalcHEP are fed into PYTHIA-6.4.21 \cite{pythia}. 
PYTHIA includes initial/final state radiations (ISR/FSR) from QED and QCD, parton showering, multiparton 
interactions and hadronisation for a realistic estimate from simulation. 
For our analysis we have used parton distribution function (PDF), CTEQ6L1 \cite{cteq6l1} from LHAPDF library \cite{LHAPDF}.
We have chosen the default factorisation ($\mu_F$) and renormalisation ($\mu_R$) scales
as set by PYTHIA.

\begin{table}[t!]
\begin{center}
\begin{tabular}{|l|l|}
\hline
\bf{Cuts used for the analysis} & \bf{Values} \\
\hline
Maximum pseudorapidity ($|\eta_\ell|$) of lepton & $ 2.5$\\
 \hline 
Minimum $p_T$ of an isolated lepton & $ 10$ GeV\\
\hline
Detector efficiency  for detecting electron ($e^\pm$) & 0.7 (70\%)\\
Detector efficiency  for detecting muon ($\mu^\pm$)   & 0.9 (90\%)\\
 \hline 
Lepton-lepton separation ($\Delta R_{\ell \ell}$) & $ \geq 0.2$ \\
 \hline
Lepton-photon separation ($\Delta R_{\ell \gamma}$) & $ \geq 0.2$ \\
\hline
Minimum hard $p_T$ cuts on $p_T$-ordered leptons & (30,30,20,20,10,10) GeV\\
\hline
Missing transverse momentum  & $\geq$ 30 GeV (for $n\ell,SS3\ell$)\\
             &         $<$ 30 GeV (for $FV4\ell$)\\  
 \hline
Lepton separated from reconstructed $jet$ ($\Delta R_{\ell j}$) & $\geq 0.4$ \\
\hline
Hadronic Activity around leptons $\frac{\sum p_{T_{hadron}}}{p_{T_{\ell}}}$ & $\leq 0.2$ \\
\hline
Electron energy smearing and muon $p_T$ resolution & $\checkmark$\\
\hline
Z veto & $|m_{\ell\ell} - M_Z| \geq 6 \Gamma_Z$ GeV \\ 
\hline
\end{tabular}
\end{center}
\caption{The cuts used to estimate SM backgrounds and signal events. For complete set of selection criteria see the text.}
\label{table:Basic_Cuts}
\end{table}

We have classified our signal space in terms of expected number of electrons and/or muons ($n \ell$ for 
n = 3 \dots 6). Two special categories for same-sign-tri-lepton ($SS3\ell$) and flavour violating four 
lepton ($FV4\ell$) processes are also created. The event selection criteria are described below: 
\begin{itemize}
 \item[I.] {\it Lepton Identification:}
\begin{itemize}
 \item[$\bullet$] Electrons and muons are identified within the pseudorapidity $|\eta_\ell| \leq 2.5$.

 \item[$\bullet$] Each of the leptons is considered to have a minimum transverse momentum of 10 GeV, i.e. ${p_T}_\ell \geq 10$ GeV.
\end{itemize}

\item[II.]{\it Lepton Efficiency:}

\begin{itemize}
 
\item[$\bullet$] Detector efficiency for detecting an isolated electron (muon) is taken as 70\% (90\%).

\end{itemize}

\item[III.]{\it  Lepton Isolation:}

\begin{itemize}

\item[$\bullet$] Two leptons are separately identified once they have a minimum separation of   
$\Delta R_{\ell \ell} \geq 0.2$, where  $\Delta R_{\ell \ell}=\sqrt{\Delta \eta^2 + \Delta \phi^2}$
is the distance in pseudorapidity ($\eta$), azimuthal angle ($\phi$) plane.

\item[$\bullet$] Leptons are separated from photons if $\Delta R_{\ell \gamma} \geq 0.2$ with all the photons having $p_{T_\gamma} > 10$ GeV. 

\item[$\bullet$] A lepton is separately identified from all reconstructed $jet$s with a minimum separation of $\Delta R_{\ell j} \geq 0.4$.

\item[$\bullet$] Isolation cuts around the hard lepton(s) should control hadronic activity. However, cleaner identification
of lepton requires hadronic activity $\frac{\sum p_{T_{hadron}}}{p_{T_{\ell}}} \leq 0.2$ around the lepton  within the cone of size 0.2.
\end{itemize}
 
\item[IV.]{\it Hard Cuts:}

\begin{itemize}
\item[$\bullet$] {\it Hard $p_T$ Cuts:}
In a multi-lepton event, we demand that the
first two hardest leptons should have minimum transverse momentum of 30 GeV, 
while that for 3rd and 4th hardest leptons are 20 GeV each.
Any additional lepton is identified with minimum transverse momentum of 10 GeV.

\item[$\bullet$]  {\it Missing Transverse Momentum:}
We demand our multi-lepton events with minimum 30 GeV of missing transverse momentum $|\not \!\! \vec{P}_T|$ 
(except in the special case of flavour violating four-lepton ($FV4\ell$) which we will discuss afterwards). 
Total missing transverse momentum $\not \!\! \vec{P}_T = - \sum_{i=1}^{N_{obj}} \vec{P}_{T_i}$ is constructed from
all reconstructed isolated objects ($N_{obj}$) such as leptons and $jet$s.
\end{itemize}

\end{itemize}

Along with the above mentioned selection criteria we have also implemented the followings in our analysis:
\begin{itemize}
 \item[$\bullet$] 
  
 The energy of electrons and $p_T$ of muons ($\mu$) are smeared according to the calorimeter resolutions.\\
 \begin{itemize}
  \item[$\ast$] 
   {\it Electron Energy Smearing:}
   We consider the smearing of the electron energy E as follows:\\
    $\sigma (E)/E = \frac{a_1}{\sqrt{E}} \oplus a_2 \oplus \frac{a_3}{E}$,\\
    where 
    \begin{center}
    \begin{tabular}{|c|c|c|c|}
     \hline
     $|\eta|$ & $a_1$ $\sqrt{\mbox{GeV}}$& $a_2$ & $a_3$ (GeV)\\
     \hline $< 1.5$ & 0.030 & 0.005 & 0.200 \\
     \hline
     $> 1.5$ & 0.055 & 0.005 & 0.600 \\
     \hline
  \end{tabular}
.
  \end{center}
 
  \item[$\ast$]
   {\it Muon $p_T$ Resolution}:
  Muon $p_T$ resolution is defined as  
  \begin{center}
  \begin{equation}
   \sigma(p_T)/p_T= \left \{ \begin{array}{lr}
                        b_1~~~~~~~ & p_T \leq 100 ~\mbox{GeV} \\
                        b_1 + b_2 ~log(p_T/100), &  p_T \geq 100 ~\mbox{GeV}.
                       \end{array} \right.
  \end{equation}
  \end{center}
  where,
      \begin{center}
    \begin{tabular}{|c|c|c|}
     \hline
     $|\eta|$ & $b_1$ & $b_2$\\
     \hline $< 1.5$ & 0.008 & 0.037  \\
     \hline
     $<2.5 ~\mbox{and}~ > 1.5$ & 0.020 & 0.050 \\
     \hline
  \end{tabular}
.
  \end{center}

 \end{itemize}

\item[$\bullet$] The $jet$s are constructed using PYCELL, cone algorithm within PYTHIA. 
To find cluster, fixed detector grid of ($100 \times 72$) assumed in ($\eta,\phi$) plane with pseudo-rapidity
$|\eta|<2.5$. With minimum threshold for $jet$ initiator $p_T$ as 1.5 GeV, a cluster can be accepted as $jet$ if minimum 
summed $E_T$ is 20 GeV within cone size 0.7. To include energy resolution of detector, energy of each cell is also smeared.

\item[$\bullet$] Z veto is implemented to reduce the SM background coming from the processes like $t\bar{t}(Z/\gamma^*)$, W(Z/$\gamma^*$), 
(Z/$\gamma^*$)(Z/$\gamma^*$). Opposite sign but same flavoured lepton pair invariant mass $m_{\ell\ell}$ must be sufficiently away 
from Z mass, such that $|m_{\ell\ell} - M_Z| \geq 6 \Gamma_Z$ GeV. However, signals remain mostly unaffected by this cut.

\end{itemize}

\noindent

We have tabulated above mentioned selection criteria in a compact form in table~\ref{table:Basic_Cuts}.

\subsection{Background Estimation}

\begin{table}[htb]
\begin{center}
\begin{tabular}{|l|r|r|r|} \hline
$\Downarrow$ \bf{processes} $\backslash$ \bf{multi-lepton channel} $\Rightarrow$  & $\bf{3 \ell}$ ($fb$)& $\bf{SS3\ell}$ ($fb$) & $\bf{4\ell}$ ($fb$)\\ \hline
$t \bar{t}$                                & $18.245$&    --                    &  --     \\    \hline
$t \bar{t} (Z / \gamma^{\star})$         & $1.121$ & $7.066 \times 10^{-4}$    & $0.069$ \\    \hline
$t \bar{t} W^\pm$                          & $0.656$ & $3.836 \times 10^{-3}$  &  --     \\    \hline
$t \bar{t} t \bar{t}$                      &   --    & $1.327 \times 10^{-4} $   &  --     \\    \hline
$t \bar{t} b \bar{b}$                      &   --    & $< 10^{-4}$              &  --     \\    \hline
$W^\pm (Z / \gamma^{\star})$             & $10.590$ &    --                    &  --     \\    \hline
$(Z/\gamma^{\star})(Z/\gamma^{\star})$ & $1.287$  &    --                    & $0.047$ \\    \hline \hline
    \bf{TOTAL}                             & $\bf{31.899}$& $\bf{4.675} \times 10^{-3}$     & $\bf{0.116}$ \\    \hline
\end{tabular}
\end{center}
\caption{Dominant SM background contributions  
to the multi-lepton channels at the LHC with $\sqrt{S}=14$ TeV
after all the cuts discussed in the subsection 5.1. K-factor for $t \bar{t}$ is taken to be 2.2. 
Blank portions represent insignificant contributions 
compared to the leading processes in that channel.
SM backgrounds for all other channels are expected to be negligible. Cross-sections are in femtobarn($fb$).}
\label{table:background}
\end{table}

Using all the cuts discussed in the previous section, 
we have estimated the SM backgrounds for different significant processes tabulated in table~\ref{table:background}. 
We have used ALPGEN-2.14  \cite{alpgen} to generate events for the following SM processes:
$t \bar{t} (Z / \gamma^{\star})$, $t \bar{t} W^\pm$, $t \bar{t} t \bar{t}$, $t \bar{t} b \bar{b}$, $W^\pm (Z / \gamma^{\star})$  
(with $0jet$) at parton level.
The ALPGEN output files are fed into PYTHIA to estimate the cross-sections for these processes. 
The SM backgrounds that emerge from the processes $t \bar{t}$, and $(Z/\gamma^{\star})(Z/\gamma^{\star})$ are estimated using 
PYTHIA. For $t\bar{t}$ process we have considered the K-factor to be 2.2 \cite{Cacciari:2008zb}.
Similar kind of analysis are performed to estimate the SM backgrounds for same-sign-tri-lepton in \cite{satya1}, 
for tri-lepton in \cite{satya2,subhadeep} and for four-lepton in \cite{satya2}.  
In passing we would like to mention that in our analysis neutral pions ($\pi^0$) are allowed to decay.\footnote{
We have noted that the neutral pion decay on/off affects the background estimation significantly due to the presence of hadronic activity cut. 
In case of pion decay one needs to implement lepton-photon isolation with a 
minimum $p_T$ cut for photons.}


\subsection{Multi-lepton Signatures} 

\begin{figure}[htb]
\includegraphics[width=5.3cm, angle=-90]{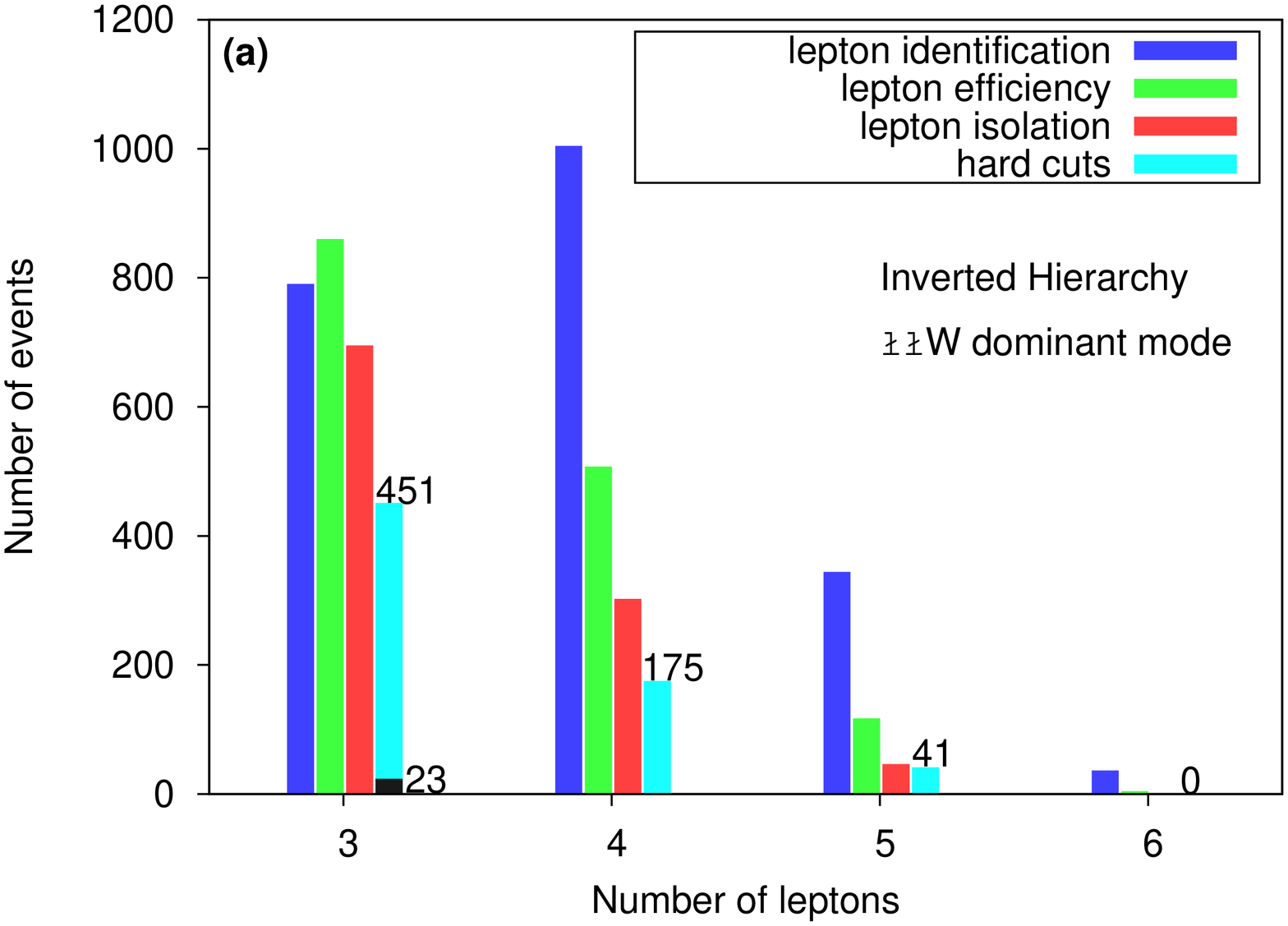}
\hspace{0.1cm}
\includegraphics[width=5.3cm, angle=-90]{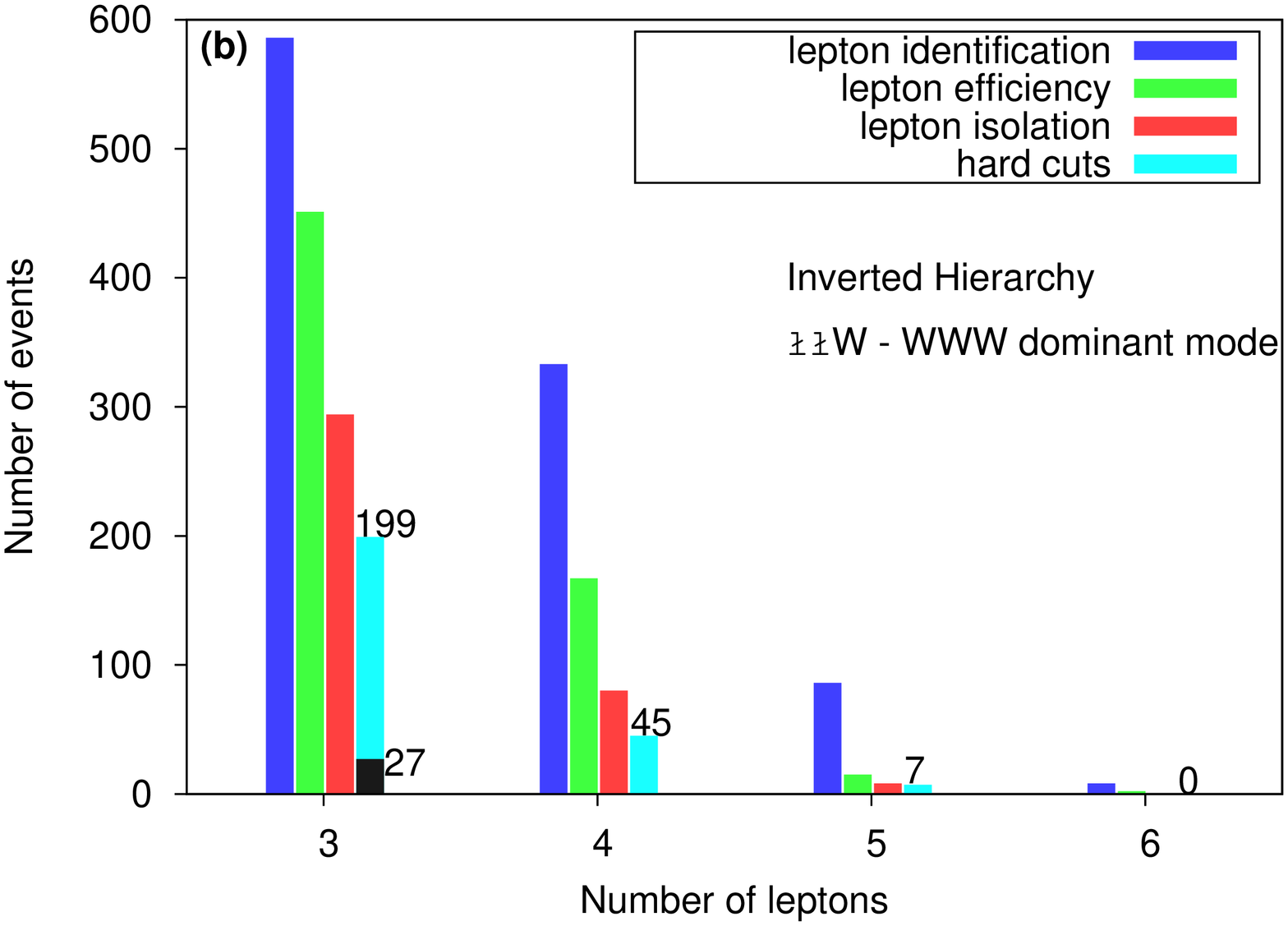}
\vspace{0.5cm}
\begin{center}
\includegraphics[width=5.3cm, angle=-90]{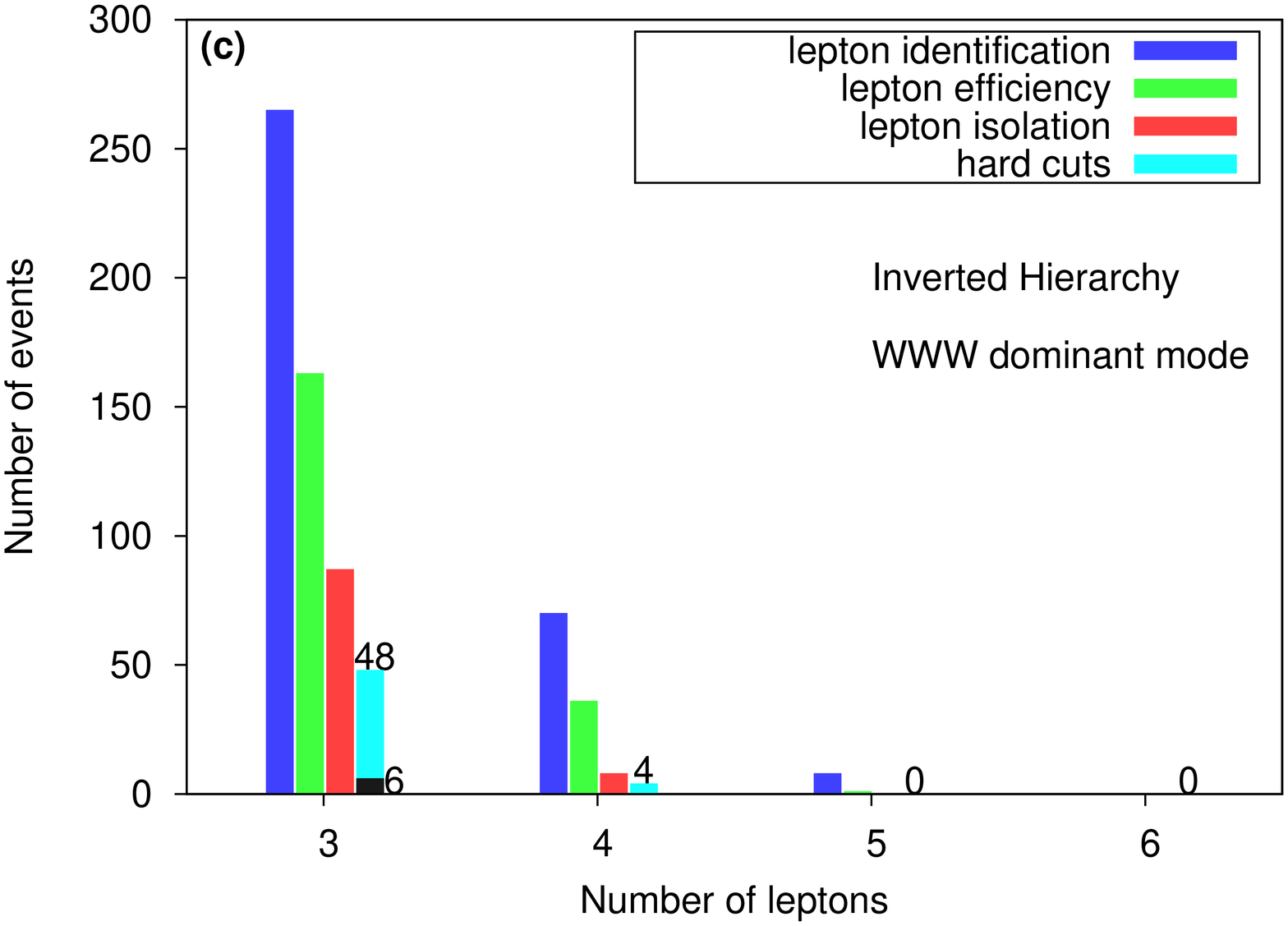}
\end{center}
\caption{The coloured histograms for a specific n-lepton signal show the number of events after implementing successive cuts
in (a) $\l{} \l{} W$, (b) $\l{} \l{} W - W W W$ and (c) $W W W$ dominant modes. The fourth column (cyan) represents 
the final multi-lepton signal events. In case of tri-lepton event the dark (black over cyan) shaded portion accounts 
for the same-sign-tri-lepton events. The final number of the respective multi-lepton events are also shown in the plots.
The number of events are computed with $M_{\Phi^0} =400$ GeV and $\Delta M = -2.8$ GeV for {\it Inverted Hierarchy} at the LHC-14
with integrated luminosity $100~fb^{-1}$.}
\label{fig:comparision_llw_IH400}
\end{figure}

\begin{figure}[htb]
\includegraphics[width=5.3cm, angle=-90]{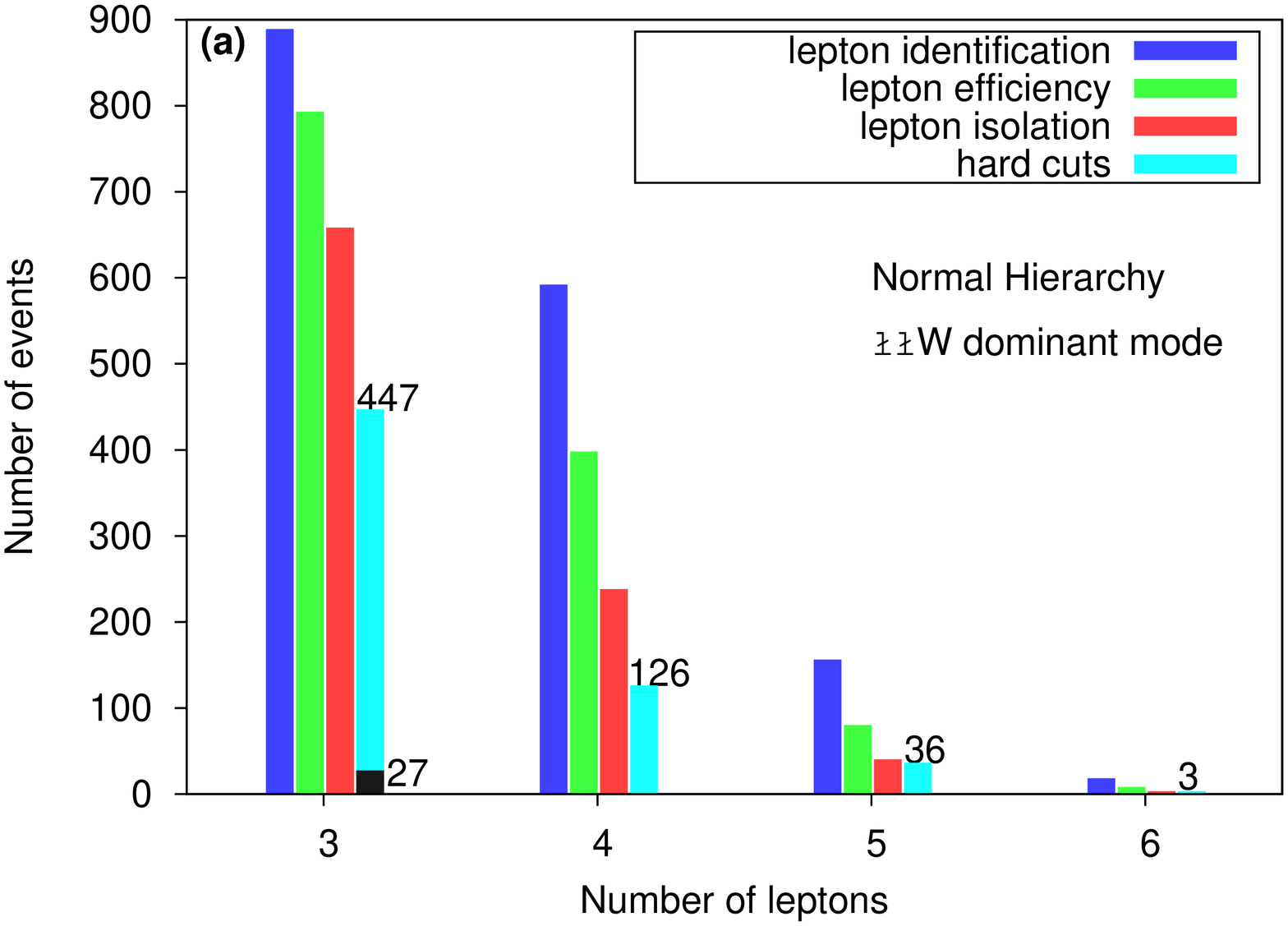}
\hspace{0.1cm}
\includegraphics[width=5.3cm, angle=-90]{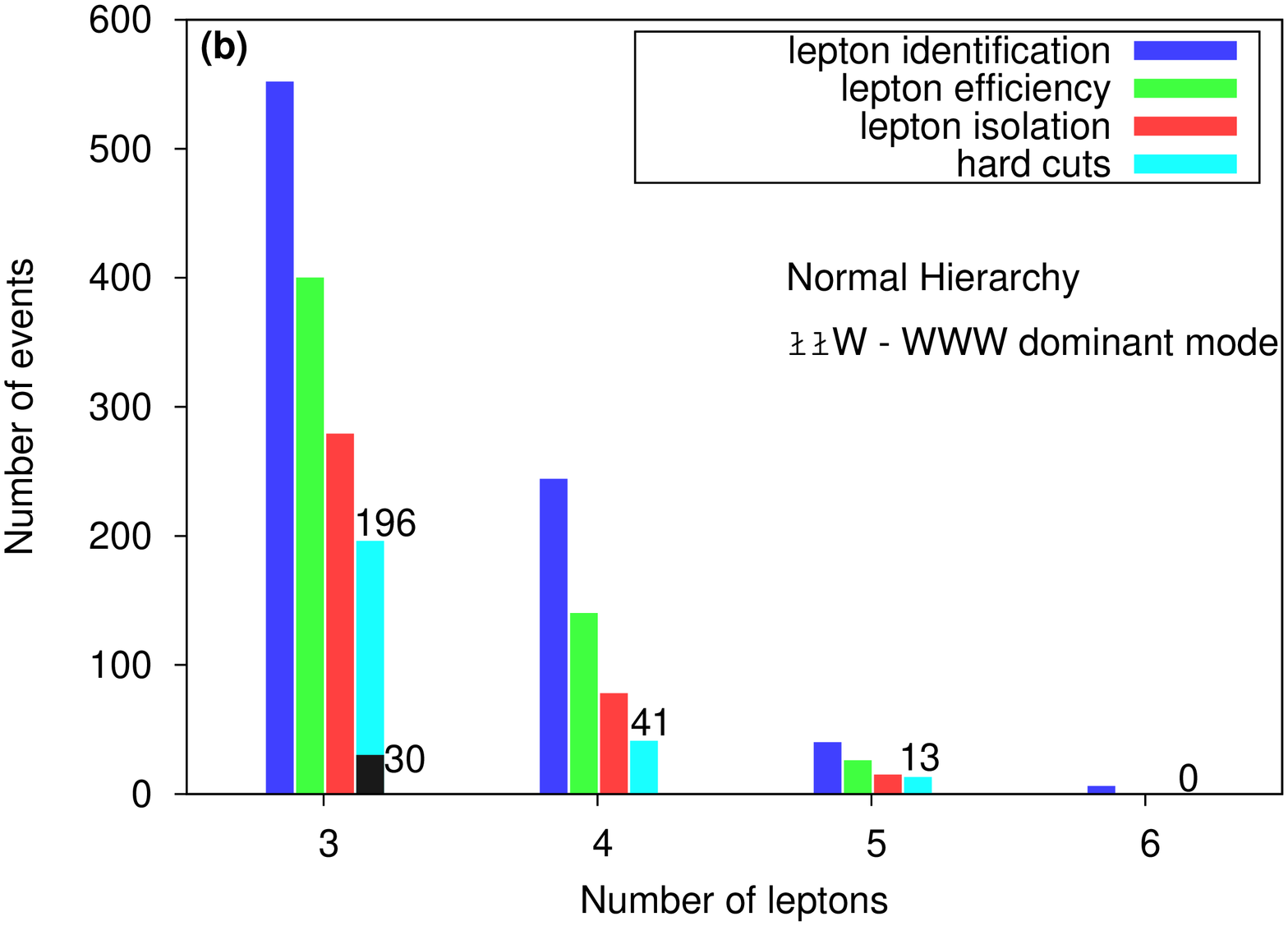}
\vspace{0.5cm}
\begin{center}
\includegraphics[width=5.3cm, angle=-90]{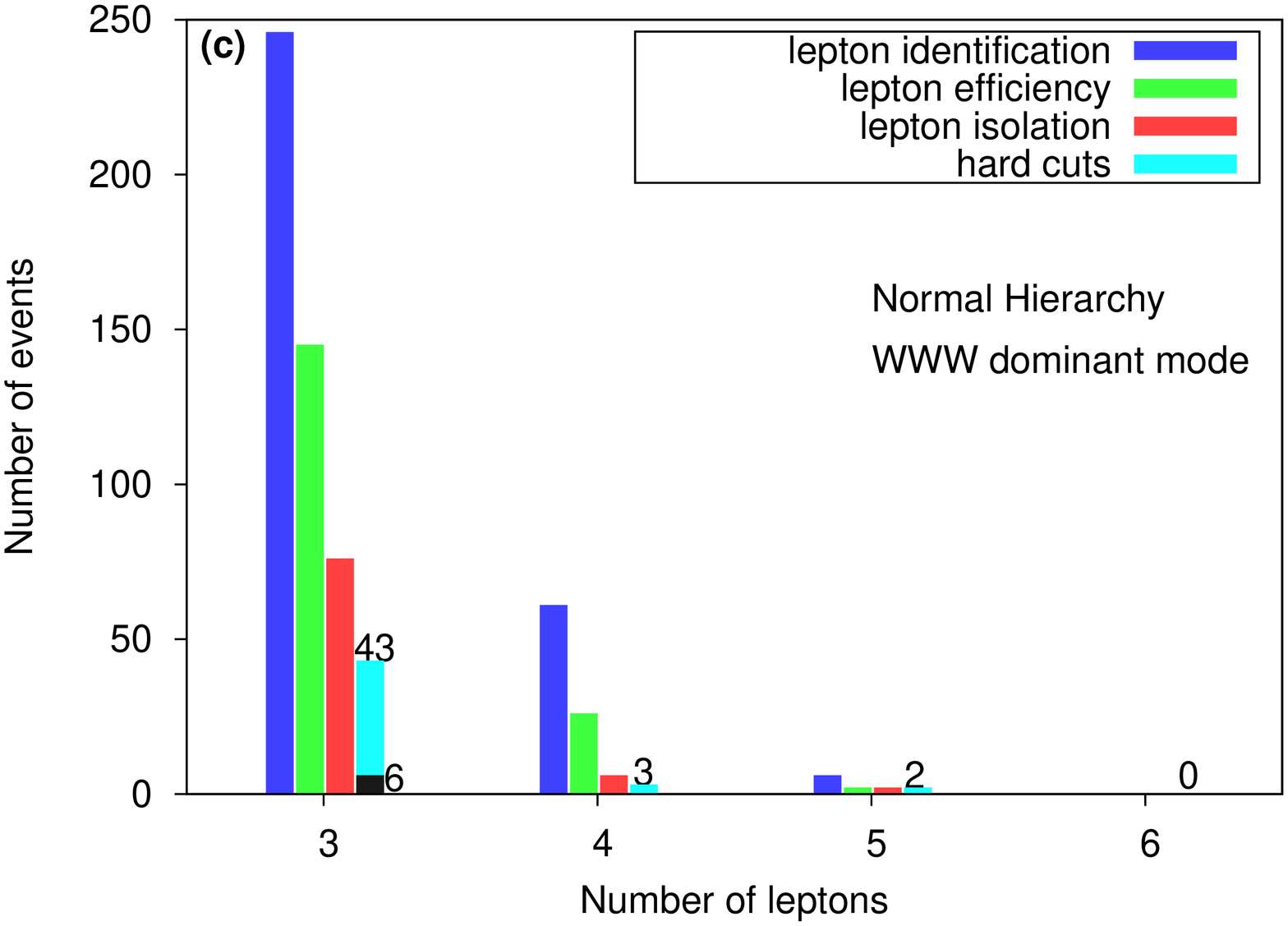}
\end{center}
\caption{The coloured histograms for a specific n-lepton signal show the number of events after implementing successive cuts
in (a) $\l{} \l{} W$, (b) $\l{} \l{} W - W W W$ and (c) $W W W$ dominant modes. The fourth column (cyan) represents 
the final multi-lepton signal events. In case of tri-lepton event the dark (black over cyan) shaded portion accounts 
for the same-sign-tri-lepton events. The final number of the respective multi-lepton events are also encoded in the plots.
The number of events are computed with $M_{\Phi^0} =400$ GeV and $\Delta M = -2.8$ GeV for {\it Normal Hierarchy} at the LHC-14
with integrated luminosity $100~fb^{-1}$.}
\label{fig:comparision_llw_NH400}
 \end{figure}

In this section we present the results for $3\ell$, same-sign-$3\ell$, $4\ell$, LFV $4\ell$, 
$5\ell$, and $6\ell$ events.  Analysis is performed with center of mass energy $\sqrt{s}=14$ TeV at the LHC with integrated luminosity $100~fb^{-1}$.
The multi-lepton signal consists of charged leptons ($e$ and/or $\mu$) + $X$, accompanied by missing transverse momentum, 
where $X$ can be associated $jet$s.
We compute the signal events for $M_{\Phi^0}=400$ GeV, $\Delta M$ = $-2.8$ GeV, and different choices of $v_\Phi$, 
mentioned in table~\ref{choice}. For each set of benchmark points 
we present the results for both Inverted Hierarchy (figure~\ref{fig:comparision_llw_IH400}) and Normal Hierarchy 
(figure~\ref{fig:comparision_llw_NH400}). 

Figure~\ref{fig:comparision_llw_IH400}(a)  
corresponds to $v_{\Phi}=5\times10^{-6}$ GeV for which $\Phi \xrightarrow{} \l{} \l{} W$ 
branching ratio is nearly 100$\%$. 
Figure~\ref{fig:comparision_llw_IH400}(b) corresponds to $v_{\Phi}=5.1\times10^{-5}$ GeV for which $\Phi \xrightarrow{} \l{} \l{} W$ and 
$\Phi \xrightarrow{} W W W$ branching ratios are $\sim 50\%$, 
while figure~\ref{fig:comparision_llw_IH400}(c) corresponds to $v_{\Phi}=0.5$ GeV for which
$\Phi \xrightarrow{} W W W$ branching ratio is $\sim 100\%$.
The x-axis represents the specific n-lepton events (n= 3,4,5, and 6) that have been  considered. 
The y-axis depicts the number of events with that particular number of leptonic events. 
We show systematically the impact of different cuts in our analysis in figure~\ref{fig:comparision_llw_IH400}.
Each coloured bar for a specific n-lepton signal shows the number of events after specific cuts. 

In each plot the first column (blue) shows the number of events after lepton identification cut.
The second column (green) is including the lepton efficiency cut. The red column at third position 
is after lepton isolation cut. Finally the fourth column (cyan) is after imposing hard $p_{T}$ and missing transverse momentum cuts.
In the fourth column only in tri-lepton events, the dark (black over cyan) shaded bar represents the same-sign-tri-lepton 
events after implementing all the above mentioned cuts.

From the plots we find that no six-lepton event survives after we impose the cuts for all three cases 
of figure~\ref{fig:comparision_llw_IH400}. There is no five-lepton event for the $WWW$ dominant mode, but for the $\l{} \l{} W$ and 
$\l{} \l{} W-WWW$ modes we get 41 and 7 events respectively. In general the number of events are more for the $\l{} \l{} W$ mode 
since the branching ratio is almost 100\% for the chosen value of $v_\Phi$. The effective leptonic BR of $WWW$ dominant mode 
for $v_\Phi=0.5$ GeV is very suppressed. Thus the number of events are suppressed. In our analysis we include 
the possible spillover from higher multiplicity events. This is noticeable for the tri-lepton events in the $\l \l W$ dominant mode 
(see figure~\ref{fig:comparision_llw_IH400}(a)).
We find 23 (27) $SS3\ell$ events in $\l \l W$ ($\l \l W- WWW$) dominant modes. For $W W W$ mode there are only 6 $SS3\ell$ events, 
see figure~\ref{fig:comparision_llw_IH400}(c).

In figure~\ref{fig:comparision_llw_NH400} we present the similar histograms corresponding
to the multi-lepton signals for NH. There are 27, 30, 6 same-sign-tri-lepton events in $\l{} \l{} W$, $\l{} \l{} W-W W W$, 
and $W W W$ dominant modes respectively. The general trend discussed in the context of figure~\ref{fig:comparision_llw_IH400} is reflected here. 
For both cases we find significant same-sign-tri-lepton, tri-lepton 
and four-lepton events over the SM background, as noted from table~\ref{table:background}. 

If we compare the total number of events for NH and IH for instance in the four-lepton channel then we see 
that they are not widely different.  
However, if one classifies these events in terms of lepton flavours then for NH  
and IH one gets relatively different number of events in each category, as can be noticed in table~\ref{table:4l_IH_NH}.
The trend in the number of events can be explained to some extent
from  neutrino mixing. The current values of mixing angles imply that the heavy states have significant amount of both $\nu_e$ and $\nu_\mu$ for IH. 
Thus one would expect somewhat similar number of events involving $e$ and $\mu$. This is reflected in table~\ref{table:4l_IH_NH}.
On the otherhand, for NH the heaviest state has relatively lower fraction of the $\nu_e$ component because of smaller value of $\theta_{13}$. 
So the number of electron events are less because the BRs are pushed in favour of more muonic events. 
This pattern is observed in all other multi-lepton channels where $\Phi \ell \ell$ vertex is involved. 
However, remember that combined results are interplay of various factors, like
$e$, $\mu$ identification efficiency and energy resolutions.

\begin{table}[htb]\centering
\begin{tabular}{|c|c|c|c|c|c|c|}
\hline
 {\bf 4$\ell$}  & $eeee$ & $eee\mu$ & $ee\mu\mu$ & $e\mu\mu\mu$  & $\mu\mu\mu\mu$ & Total events\\
\hline
IH  & 14 & 47 & 69 & 29 & 16 & 175 \\
\hline
NH  & 1 & 1 & 23 & 40 & 61 & 126 \\
\hline
\end{tabular}
\caption{ Neutrino mass hierarchy dependency in four-lepton signal in \l{}\l{}W dominant region.}
\label{table:4l_IH_NH}
\end{table} 

\subsection{Lepton Flavour Violating Signal}  
\label{sec:LFV}

The multi-lepton events obtained in this model can be of mixed flavours and one can study 
the charged lepton flavour violation at the LHC. 
The interaction which is mainly responsible for this signal is the effective vertex, 
$\Phi^{\pm\pm} \ell^\mp_i \ell^\mp_j$, discussed in 
appendix~\ref{appendix_feynman_rules_effective_vertex}.
This is proportional to the light neutrino mass matrix elements $m_{\nu_{ij}}$, with $i,j=e,\mu$.  
Of special importance in this respect are the four-lepton signals. These are not 
accompanied by any neutrino in the final
state and hence the flavour of all the final state leptons can be ascertained. 
Note that these type of signals are originated from the 
inclusive pair productions of doubly-charged scalars. 
Based on the lepton flavours these  signatures can be categorised into two classes:
i) $p p \rightarrow \ell_i^{+} \ell_i^{+} \ell_j^{-} \ell_j^{-} ~+~ X$, and
ii) $p p \rightarrow \ell_i^{\pm} \ell_j^{\pm} \ell_j^{\mp} \ell_j^{\mp} ~+~ X$,
with $\ell_i \neq \ell_j=e, \mu$. The former final state emerges when each of the $\Phi^{\pm \pm}$ 
decays into same flavours of charged leptons, but the latter arises if one of the $\Phi^{\pm \pm}$
decays into different flavours. Our parton level signals consist of four-leptons + $X$ with no missing transverse momentum. 
Hence, we demand very small $|\not \!\! \vec{P}_T|$ (< 30 GeV) in event selection. 
We present the results for the LFV four-lepton signals in table~\ref{4lfv-signal}.
For our study we consider LFV signal in the $\l{} \l{} W$ and $\l{} \l{} W - W W W$ dominant regions. In the 
$W W W$ dominant region since the decay of $W$'s leads to large missing energy coming from neutrinos, 
the lepton flavour violating nature of the final state cannot be determined. 
Hence, this region does not contribute to our signal.

From table~\ref{4lfv-signal} we see that the number of events for NH and IH are different. 
This difference is more pronunced for the events of the first class which are the $ee\mu\mu$ type of events.
In this case, the number of events are much less for NH because $(m_{\nu})_{ee}^{NH}<(m_{\nu})_{ee}^{IH}$.
For the second class there are two type of events -- $eee\mu$ and $e\mu\mu\mu$. 
However, because of the higher detection efficiency  of the muons the latter type of events give the dominant 
contribution. These are goverened by the elements $(m_{\nu})_{e\mu}$ and $(m_{\nu})_{\mu\mu}$. For our choice of 
parameters $(m_\nu)_{\mu\mu}^{NH} \approx (m_\nu)_{\mu\mu}^{IH}$ and $(m_\nu)_{e\mu}^{NH} > (m_\nu)_{e\mu}^{IH}$. 
Thus we get more number of events for NH.
In the $\l{} \l{} W- WWW$ dominant region, same trend can be observed though the number of events are significantly 
smaller since $W W W$ channel does not contribute to this signal.  

\begin{table}[htb]\centering
\begin{tabular}{|c|c|c|c|c|}
\hline      
Dominant & Hierarchy  & \# of events &  \# of events & Total\\
 decay region &  & ($\ell_i^{+} \ell_i^{+} \ell_j^{-} \ell_j^{-} +X$)  & 
 ($\ell_i^{+} \ell_j^{+} \ell_j^{-} \ell_j^{-} + X$) & \\
\hline
$\l{} \l{} W $& IH & 22  & 4 & 26 \\
\hline
$\l{} \l{} W$& NH & 0  & 9 &  9 \\
\hline
$\l{} \l{} W - W W W$& IH & 4 & 0 & 4 \\
\hline
$\l{} \l{} W - W W W$& NH & 0 & 3 & 3 \\
\hline
\end{tabular}
\caption{Lepton flavour violating four-lepton signals at different dominant decay regions for normal and inverted hierarchies at the LHC . 
The signal events are computed using parameters $M_{\Phi^0}= 400$ GeV, $\Delta{M}  = -2.8$ GeV, $|\not \!\! \vec{P}_T|<30$ GeV. }
\label{4lfv-signal}
\end{table}


\subsection{New Contributions to $H \rightarrow \gamma \gamma$} 
\label{sec:h2aa}

\begin{figure}[htb]
\centerline{\includegraphics[scale=0.65]{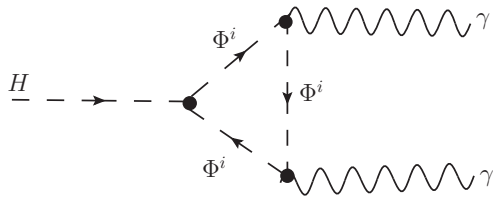}}
\caption{\small New diagrams that contribute in Higgs to di-photon decay through charged components of $\Phi$.
Here $\Phi^i$ represents singly, doubly and triply-charged scalars.}
\label{figure:h2gammagamma}
\end{figure}

The charged scalars in this model couple to both neutral Higgs as well as the
photon. Thus, they lead to added contributions to the Higgs to di-photon ($\gamma \gamma$) process. 
The dominant contribution comes from the 
diagram shown in figure~\ref{figure:h2gammagamma}.
The relevant part of the Lagrangian reads as 
\beqa
\mathcal{L}_{H\gamma \gamma} = 
(y_{3} \Phi^{+++} \Phi^{---} + y_{2} \Phi^{++} \Phi^{--} + y_{1} \Phi^{+} \Phi^{-}) H v,
\eeqa
where $y_3=2\lambda_3-3\lambda_4/2,~ y_2=2\lambda_3-\lambda_4/2,~ y_1=2\lambda_3 +\lambda_4/2.$

\begin{figure}
\centerline{\includegraphics[width=8cm, height= 8cm]{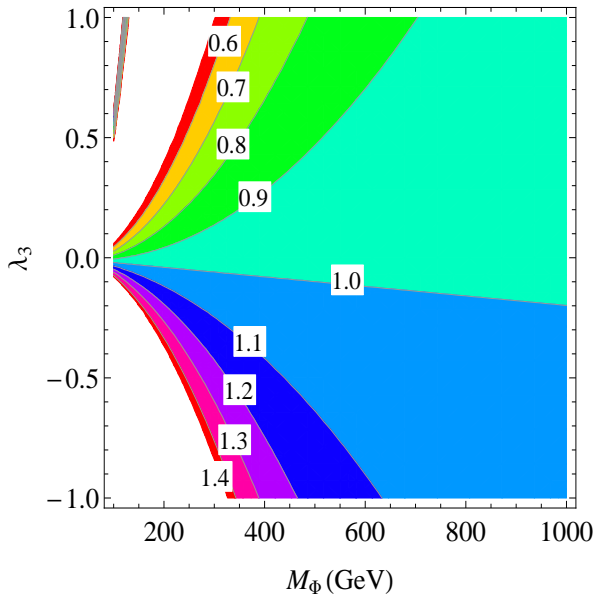}}
\caption{\small Iso-contour of $R_{\gamma \gamma}$ in the $M_{\Phi}-\lambda_3$ plane.}
\label{figure:rgammagamma}
\end{figure}

In figure~\ref{figure:rgammagamma} we plot the iso-contours of $R_{\gamma \gamma}$ in the $M_{\Phi}-\lambda_3$ plane, where 
$R_{\gamma \gamma}$ is the ratio of the partial decay widths of Higgs to di-photon for new model and that for the 
SM. To minimize the number of free parameter we have varied $\lambda_3$ within perturbative limit and $\lambda_4$ 
is reconstructed from the following relation $\Delta M^2 = \frac{\lambda_4}{2} v^2$, with $\Delta M=-2.8$ GeV.
We note that the Higgs to di-photon rate in this model can be larger (smaller) compared to the SM for $\lambda_3<0$ ($\lambda_3>0$) 
for values of $M_{\Phi^0}$ in the LHC accessible range.
Note that the multi-lepton signals that we consider does not depend on 
the parameter $\lambda_3$. Since in the model under consideration the vector-like fermions 
are heavy their contribution to the di-photon decay rate is suppressed.


\section{Conclusion}\label{sec:conclusion}

In this paper we consider a model which can  generate  neutrino masses through
an effective dimension-7 operator. 
This requires the presence of an isospin 3/2 scalar
and a pair of $Y=2$ vector-like $SU(2)$ triplet fermions.
The neutral fermion mass matrix is of the linear seesaw form
and one can get light neutrinos of mass in the right ballpark 
even if the new particles  are at $\mathcal{O}$(TeV)  scale. 
We choose the scalar quadruplet to be of mass lower than TeV such that the charged scalars belonging to this can be  
pair produced at the LHC.  
Subsequent decays of these scalars to leptons or W-bosons 
and further decays of W-boson produce multi-lepton final states. 
We study the tri-lepton, same-sign tri-lepton, four-lepton, five-lepton 
and six-lepton signals in this model
at the LHC at $\sqrt{s}=14$ TeV with integrated luminosity
$\int \mathcal{L} dt = 100 ~fb^{-1}$. 
A noteworthy feature in this model is the presence of the effective vertex
$\Phi^{\pm \pm} \ell^{\mp} \ell^\mp$ which  facilitates the same-sign-tri-lepton 
events for which the SM background is not significant and hence they 
can herald new physics beyond SM. 
Further more since this vertex depends on the neutrino mass
matrix elements, it induces a dependence on the neutrino mass hierarchy in the observed signal. 
We do a realistic simulation using CalcHEP and  PYTHIA 
incorporating appropriate cuts.
We also estimate the SM background using ALPGEN.
We choose the parameters of the model to cover the different dominant decay modes of the charged scalars. 
 Among the events studied,
the 6 lepton events do not survive the cuts for most of the benchmark points. 
For the other multi-lepton events significant excess over the SM background 
can be observed. Another hallmark of this model is the possibility  of obtaining
flavour violating four-lepton signal. We investigate this
option in the context of the LHC and find significant number of events. 
We estimate the additional contribution
to the $H \rightarrow \gamma \gamma$ rate in this model
and delineate the parameter space in which this rate can deviate from the SM value.

In conclusion, the model considered in this paper is
phenomenologically rich, can generate small neutrino mass consistent with data
and can also be probed at the LHC through the multi-lepton signatures.

\section{Acknowledgements} Authors would like to acknowledge K.S. Babu and S. Roy for useful comments.
GB and SG wish to thank  S. Khan for useful discussions. 
JC thanks S. Biswas, S. Mondal and T. Mondal for technical helps.
GB wants to thank K. Patel for useful discussions. 

\newpage


\appendix


\section{Quadruplet Scalar Kinetic Term}
\label{appendix_scalar}

The quadruplet scalar kinetic term in eq.~\ref{lphi} reads as:
\begin{equation}
 \mathcal{L} = (D^{\mu} \Phi )^{\dagger} (D_{\mu} \Phi).
\label{PhiGBterm}
\end{equation}

The Feynman rules for the interaction of the $\Phi$ field with gauge bosons can be obtained 
from the covariant derivative 
\begin{equation}
D_{\mu} \Phi = \left(\partial_{\mu} -i g \vec{T} {\bf .}  \vec{W}_{\mu} - i g' \frac{Y}{2}B_{\mu}\right) \Phi .
\label{PhiGB}
\end{equation}
Since the $\Phi$ belongs to the isospin-3/2 representation of 
$SU(2)$, the generators $T_{a}$ can be expressed as, 
\beqa
T_1=
\begin{array}{cccc}
\begin{pmatrix}
0 & \sqrt{3}/2 & 0 & 0 \\
\sqrt{3}/2 & 0 & 1 & 0 \\
0 & 1 & 0  & \sqrt{3}/2 \\
0 & 0 & \sqrt{3}/2 & 0 
\end{pmatrix},
\end{array}
~~
T_2=
\begin{array}{cccc}
\begin{pmatrix}
0 & -i \sqrt{3}/2 & 0 & 0 \\
i \sqrt{3}/2 & 0 & -i & 0 \\
0 & i & 0  & -i \sqrt{3}/2 \\
0 & 0 & i \sqrt{3}/2 & 0 
\end{pmatrix},
\end{array} 
\label{Ti} 
\eeqa 

\beq
T_3 = diag(3/2,1/2,-1/2,-3/2) \nonumber.
\eeq


\section{Yukawa Interactions of Fermion Triplets}
\label{appendix_feynman_rules_Yukawa}

The Yukawa Lagrangian for the $\Sigma$ field is given as:
\beqa
{\cal L}_{Y}  
= Y_i \overline{{l_{iL}}^{C}_a} H^{b} {\Sigma_L}_{a^{'}b} \epsilon^{aa^{'}}
+ {Y_i^\prime} {\overline{\Sigma_R}}^{ab} \Phi_{abc} {l_{iL}}_{c^{'}} \epsilon^{cc^{'}}
+ h.c., 
\label{Lmass}
\eeqa
where $a,b$ etc. are $SU(2)$ indices. 

The components of $\Sigma_{L,R}$ are:
\begin{eqnarray}
{\Sigma}_{11} = \Sigma^{++}, {\Sigma}_{12} = \frac{1}{\sqrt{2}} {\Sigma}^{+},
{\Sigma}_{22} = {\Sigma}^{0}. \nonumber 
\end{eqnarray}

The components of quadruplet scalar field $\Phi$ are:
\begin{eqnarray}
\Phi_{111} = \Phi^{+++}, && \Phi_{112} = \frac{1}{\sqrt{3}} \Phi^{++},  \\
\Phi_{122} = \frac{1}{\sqrt{3}} \Phi^{+},  && \Phi_{222} = \Phi^{0}. \nonumber 
\end{eqnarray}

In the form of component fields the terms in eq.~\ref{Lmass} can be written as:
\beqa
 \overline{{l_{iL}}^{C}}H^{*} \Sigma_L  
= \overline{{\nu_{iL}}^{C}} H^{-} \frac{1}{\sqrt{2}}{\Sigma_L}^{+} + 
 \overline{{\nu_{iL}}^{C}} H^{0} {\Sigma_L}^{0} - 
 \overline{{l_{iL}^{-}}^{C}} H^{-}{\Sigma_L}^{++}
- \overline{{l_{iL}^{-}}^{C}} H^{0} \frac{1}{\sqrt{2}}{\Sigma_L}^{+}, 
\label{LHSigma} 
\eeqa
\beqa
{\overline{\Sigma_R}}^{ab} \Phi_{abc} {l_{iL}}_{c^{'}} \epsilon^{cc^{'}}
&  = &   
\overline{{\Sigma_R}^{++}} \Phi^{+++} {l_{iL}^{-}} + 2\frac{1}{\sqrt{2}} \overline{{\Sigma_R}^{+}} 
\frac{1}{\sqrt{3}}\Phi^{++} {l_{iL}^{-}} - 2\frac{1}{\sqrt{2}} \overline{{\Sigma_R}^{+}} \frac{1}{\sqrt{3}}\Phi^{+} \nu_{iL} \nonumber \\
&  & - \overline{{\Sigma_R}^{++}} \frac{1}{\sqrt{3}}\Phi^{++} \nu_{iL} - \overline{{\Sigma_R}^{0}} \Phi^{0} \nu_{iL} + 
\overline{{\Sigma_R}^{0}} \frac{1}{\sqrt{3}}\Phi^{+} {l_{iL}^{-}}. 
\label{LphiSigma} 
\eeqa


\section{Neutrino Mass through Dimension-7 Effective Vertex}
\label{appendix_feynman_rules_effective_mass}

\begin{figure} 
\includegraphics[scale=0.65]{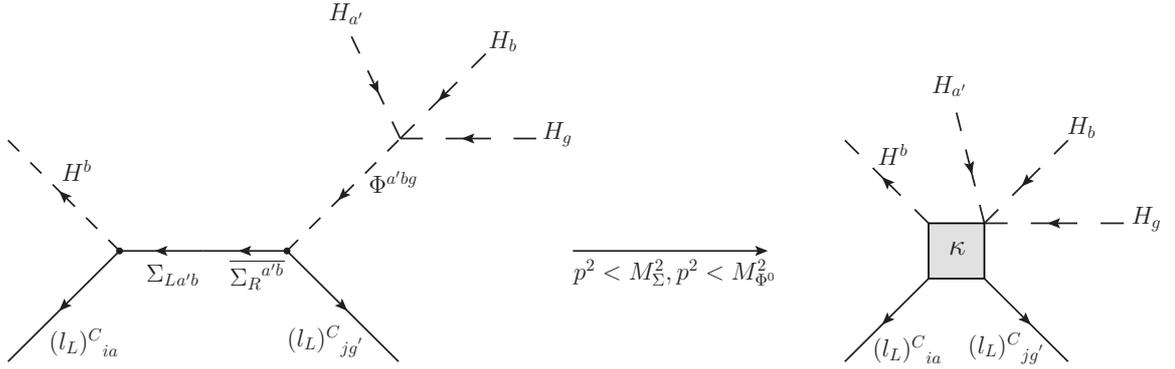}
\caption{The tree-level diagram for the generation of the dimension-7 
effective vertex is shown on the left-hand side. The right-hand figure shows the 
effective vertex in the low energy approximation.} 
\label{Dimension_7_operator_effectiveFT}
\end{figure}

In this sub-section we discuss the tree level diagram
which gives rise to the dimension-7 effective operator, see figure~\ref{Dimension_7_operator_effectiveFT}. 

The tree level diagram can be evaluated in the small momentum 
transfer limit as, 
\begin{eqnarray}
M_{ij}  & =  & P_{L} {(Y_i Y_j^{'})} \frac{1}{\slashed{p}-M_{\Sigma}} \frac{1}{p^2 - M^2_{\Phi^0}}\lambda_5 \varepsilon_{aa^{'}} 
\varepsilon_{gg^{'}}P_{L} + \quad i \leftrightarrow j  \nonumber \\
& & 
\xrightarrow[p^{2} < M^2_{\Sigma} , p^{2} < M^2_{\Phi^0}]{} {(Y_i Y_j^{'})} \frac{1}{M_{\Sigma}} \frac{1}{M^2_{\Phi^0}}\lambda_5 
\varepsilon_{aa^{'}} \varepsilon_{gg^{'}}P_{L} + \quad i \leftrightarrow j .\quad \quad \quad \quad \quad
\end{eqnarray}


The diagram on the right-side can be evaluated as,

\beqa
{\cal L}_\kappa & = & \kappa_{ij}
\left( {\overline{l_L^C}}^i \sigma^{\alpha} \varepsilon H \right)
\left( H^T \sigma^{\alpha} \varepsilon l_L^j \right)(H^{\dagger}H) + {\rm h.c.} ,
\\
&=& - \kappa_{ij} \left( {\overline{l_L^C}}^i_a
H_{a^{'}}  l_{Lg^{'}}^j H^{b}H_{b}H_{g} \right) \varepsilon_{aa^{'}} \varepsilon_{gg^{'}} + {\rm h.c.} \; \; ,
\label{L-kappa}
\eeqa
which leads to
\begin{equation}
M_{ij}= -\kappa_{ij} \epsilon_{aa^{'}} \epsilon_{gg^{'}}P_{L}.
\label{dim7_amplitude}
\end{equation}

Tree level matching then gives, 
\begin{equation}
\kappa_{ij} = -\frac{(Y_i Y_j^{'} + Y_i^{'} Y_j) \lambda_5 }{M_{\Sigma}{M^2_{\Phi^0}}}.
\label{dim7_kappa}
\end{equation}

Therefore, the neutrino mass after symmetry breaking is  \\
\begin{equation}
 m_{\nu_{ij}} =  -\frac{(Y_i Y_j^{'} + Y_i^{'} Y_j) \lambda_5 }{M_{\Sigma}{M^2_{\Phi^0}}} v^{4}.
\end{equation}


\section{Lepton Flavour Violating Effective Vertex}
\label{appendix_feynman_rules_effective_vertex}

\begin{figure}[htb]
\begin{center}
 \includegraphics{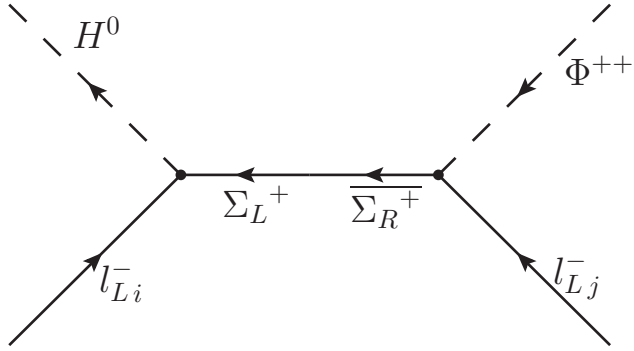}
 \caption{Effective vertex of $\Phi^{++}\ell^{-}_i\ell^{-}_j$ coupling.}
 \label{effective_vertices}
 \end{center}
\end{figure}

There is an interesting lepton number violating vertex which arises in this
model from the diagram  in figure~\ref{effective_vertices}.

 In the limit of small momentum transfer, integrating out the 
heavy fields $\Sigma$, this diagram gives rise to an effective  
$\Phi^{++}l^{-}_il^{-}_j$ vertex, which after the 
$H^0$ field gets VEV gives $\frac{m_{\nu_{ij}}} {2\sqrt{3}v_{\Phi}}$.
The singly charged  and neutral scalar ($\Phi^{\pm}$, $\Phi^{0}$) 
can also have similar effective vertex and can decay to a lepton and a neutrino 
or two neutrinos.

\section{Feynman Rules}
\label{feynmanrules}
In this section we tabulate the  Feynman rules required in the 
calculation, involving the 
additional particles in the model -- namely the 
the isospin 3/2 scalar and the vector-like triplet fermions. 
We also tabulate the Feynman rule corresponding to the dimension-7 effective
operator obtained by integrating out the triplet fermions and the isospin 3/2 scalars. 
The arrows on the fermion lines indicate the direction
of the lepton number flow.

\newpage
\begin{itemize} 
\item
{\bf{Feynman rules relevant for production and detection of $\Phi$}
}

The interactions from the expansion of eq.~\ref{PhiGBterm} give rise to the following vertices used in our calculations.
The last one comes from the diagram  discussed in appendix~\ref{appendix_feynman_rules_effective_vertex} and depends on the 
effective neutrino mass. 

\begin{figure}[htb]
\includegraphics[width=6.5cm,angle=0]{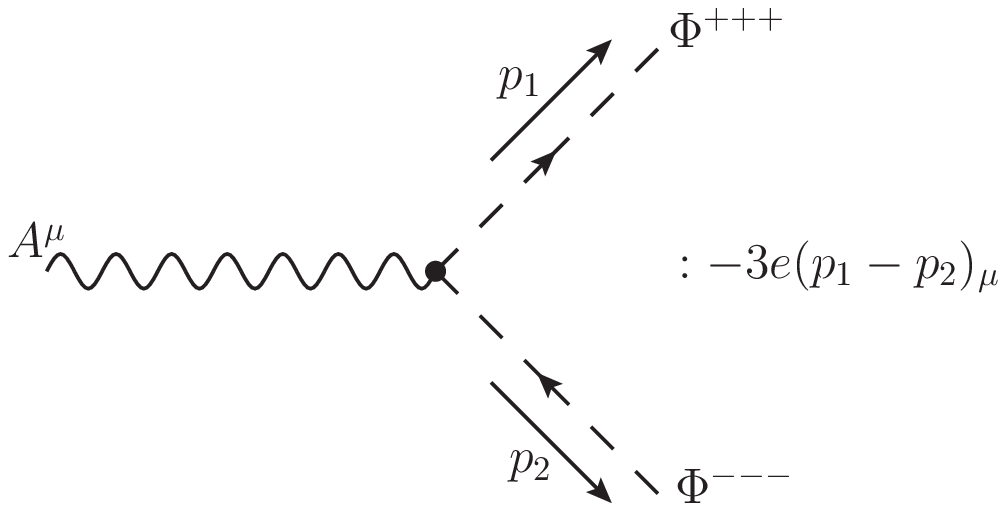}
\; \; \; \; 
\includegraphics[width=7.5cm,angle=0]{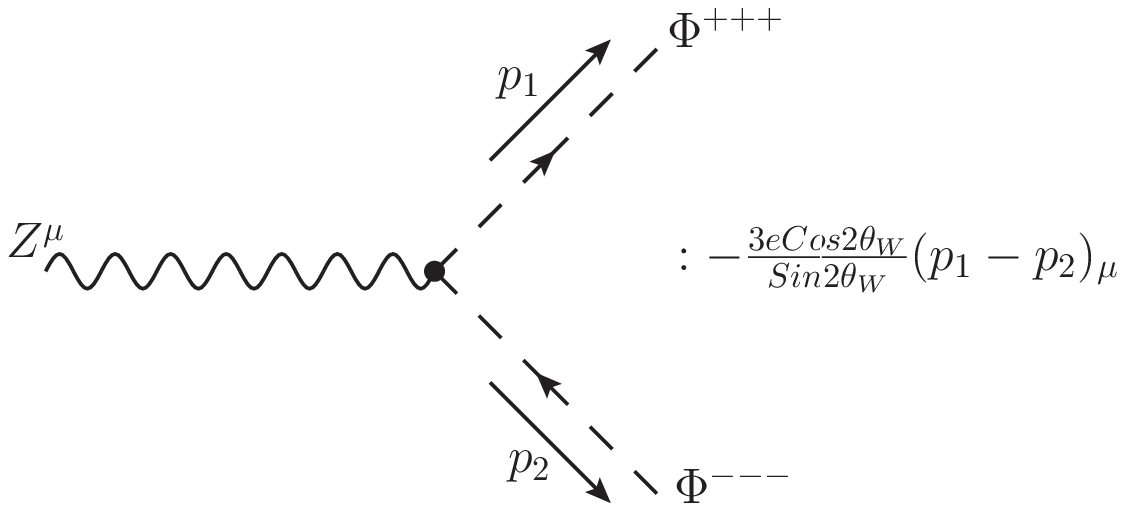}

\includegraphics[width=6.5cm,angle=0]{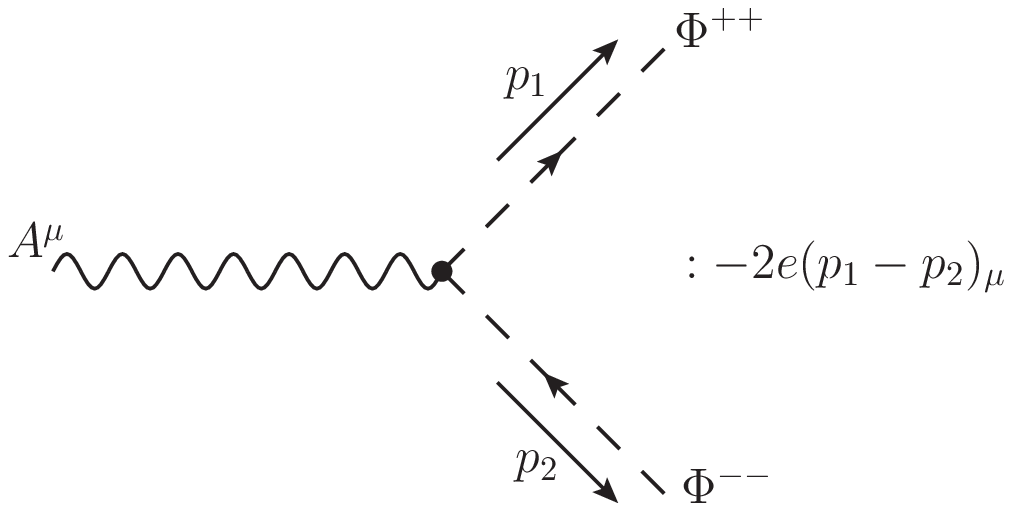}
\; \; \; \;
\includegraphics[width=7.5cm,angle=0]{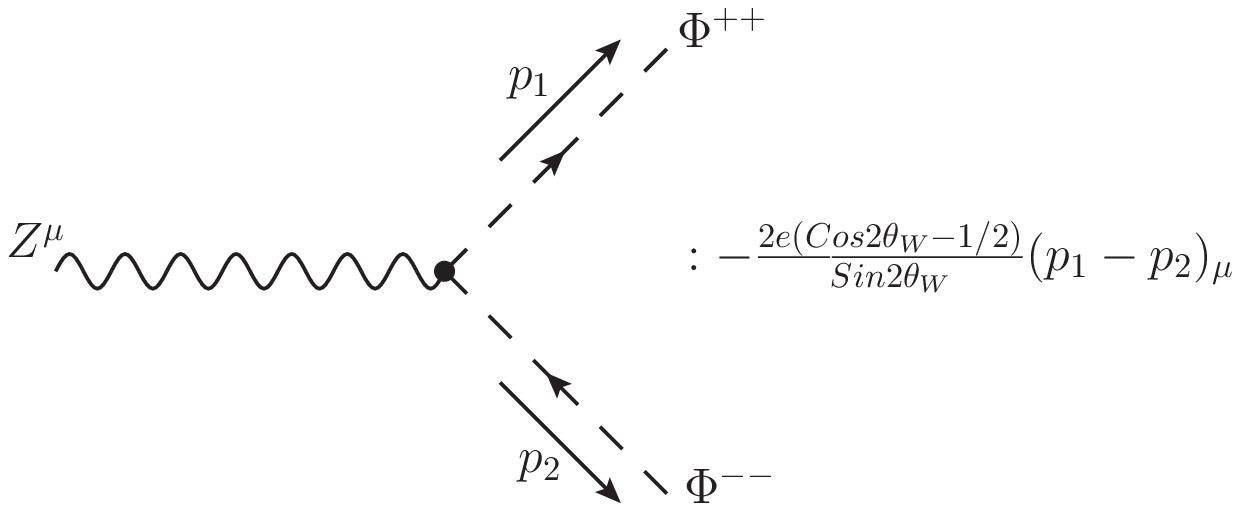}

\includegraphics[width=6.5cm,angle=0]{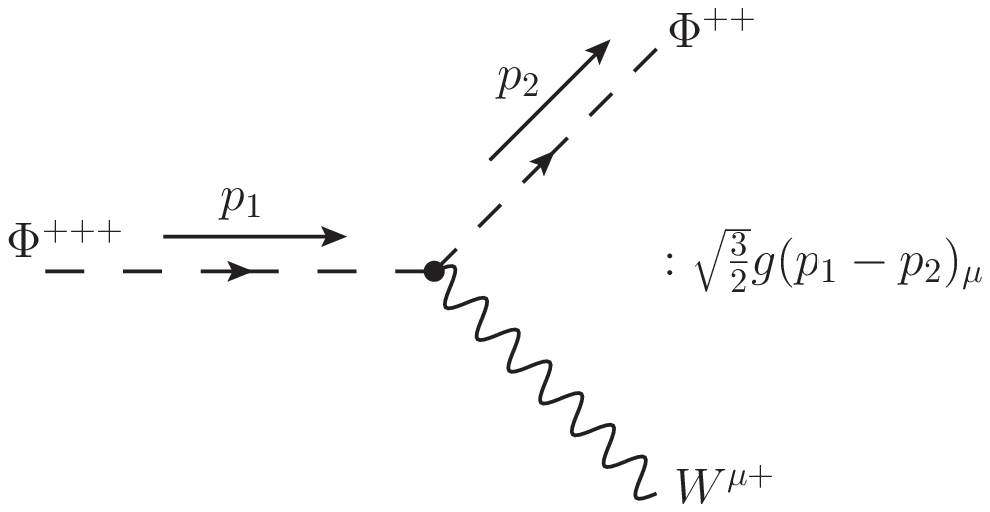}
\; \; \; \;
\includegraphics[width=5.5cm,angle=0]{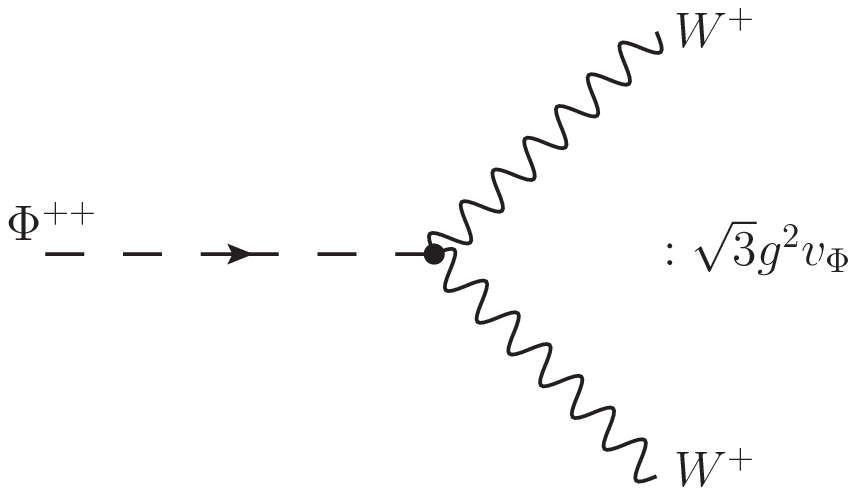}
\label{coupling}
\end{figure}
\begin{figure}[h]
\includegraphics[width=5.5cm,angle=0]{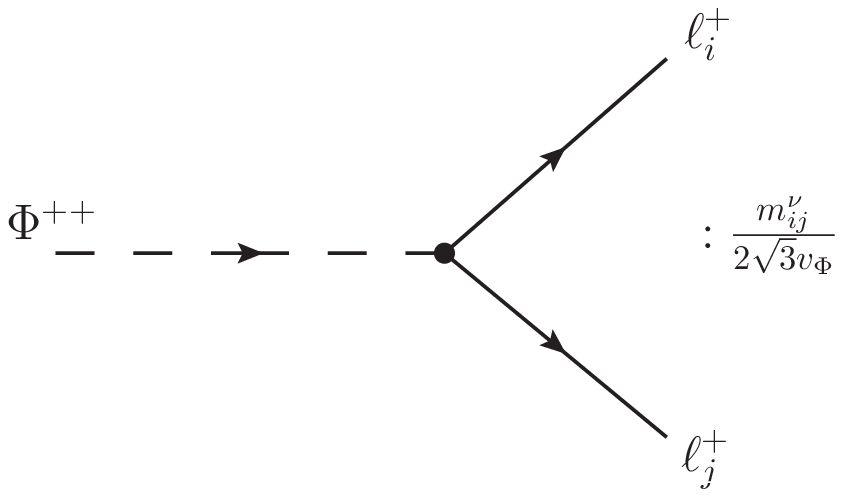}
\caption{Feynman rules for the production and decays of charged scalars.}
\label{coupling_eff}
\end{figure}


\newpage
\item{\bf{The Feynman rules for the Yukawa Interactions of $\Sigma$}}

The Yukawa Interactions of $\Sigma$ are calculated in section~\ref{appendix_feynman_rules_Yukawa}.
The vertex factors are extracted from eq.~\ref{Lmass}.

\begin{figure}[htb]
\includegraphics[width=6cm,angle=0]{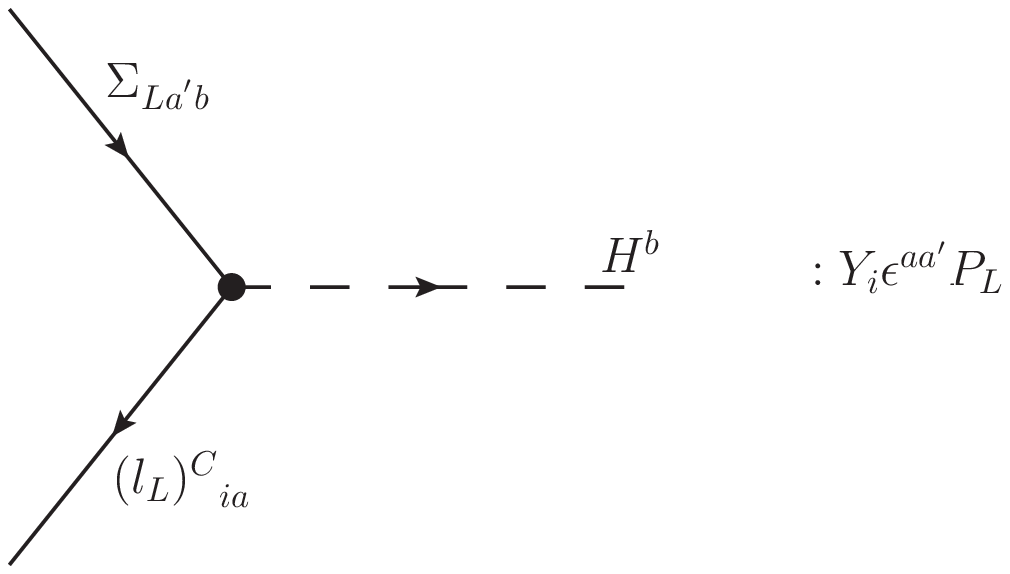}
\; \; \; \;
\includegraphics[width=6cm,angle=0]{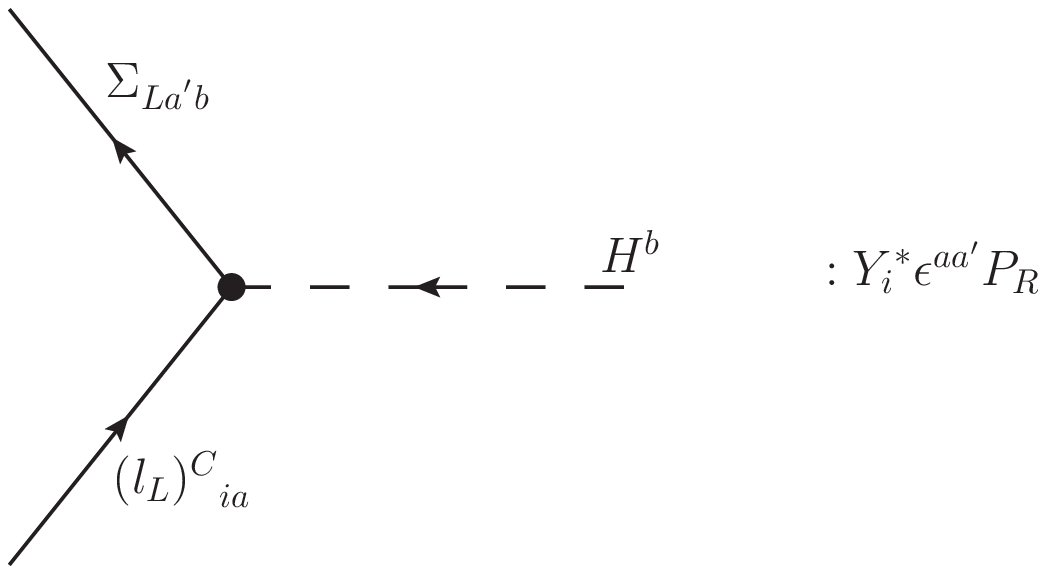}
\; \; \; \;

\includegraphics[width=6cm,angle=0]{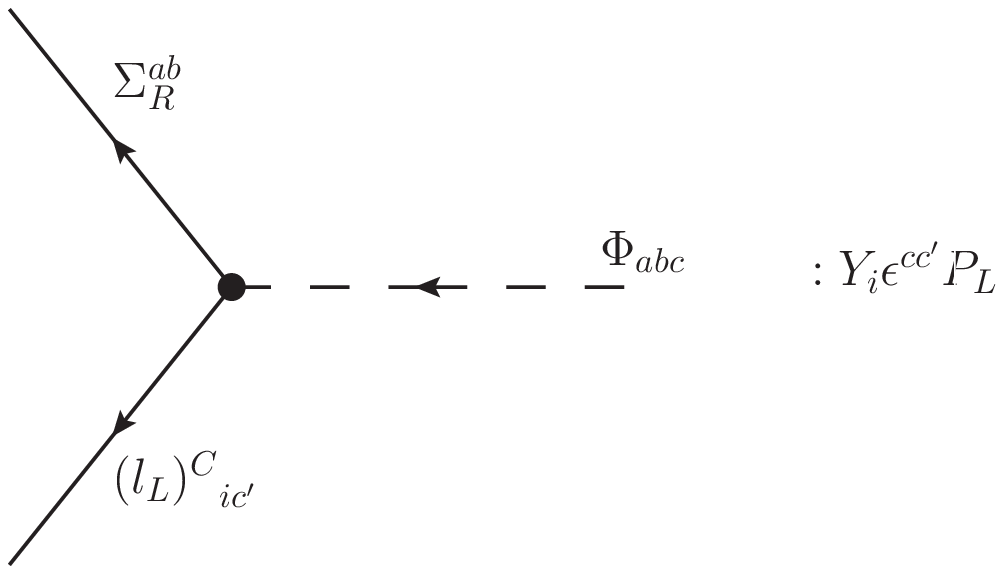}
\; \; \; \;
\includegraphics[width=6cm,angle=0]{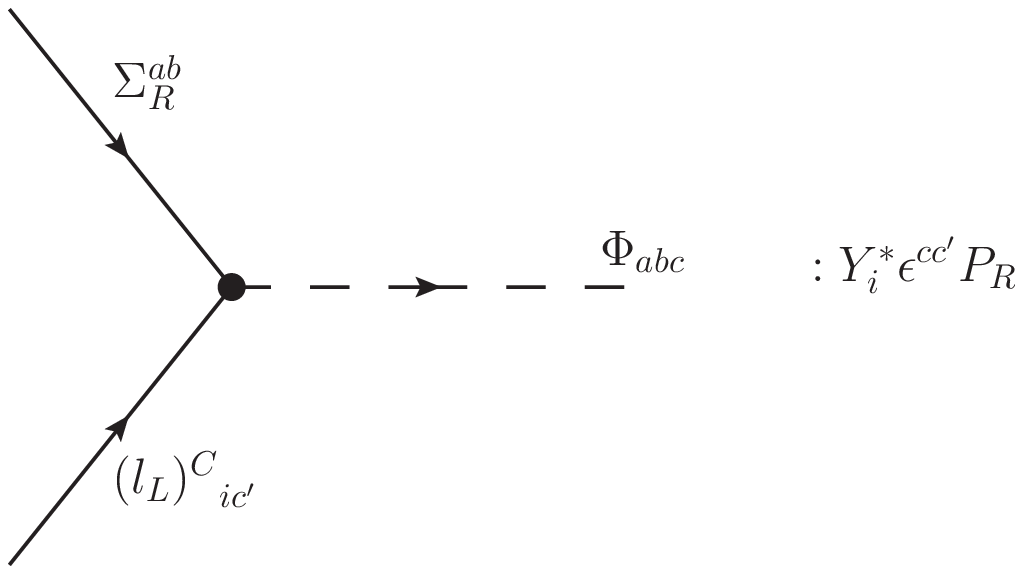}

\; \; \; \;
\label{coupling_Yukawa}
\caption{Feynman rules for Yukawas in the Lagrangian.}
\end{figure}

\item{\bf{The Feynman rules for effective vertex $\kappa$}}

The effective vertex $\kappa$ is derived in eqs.~\ref{dim7_amplitude},~\ref{dim7_kappa}.

\begin{figure}[htb]
\begin{center}
\includegraphics{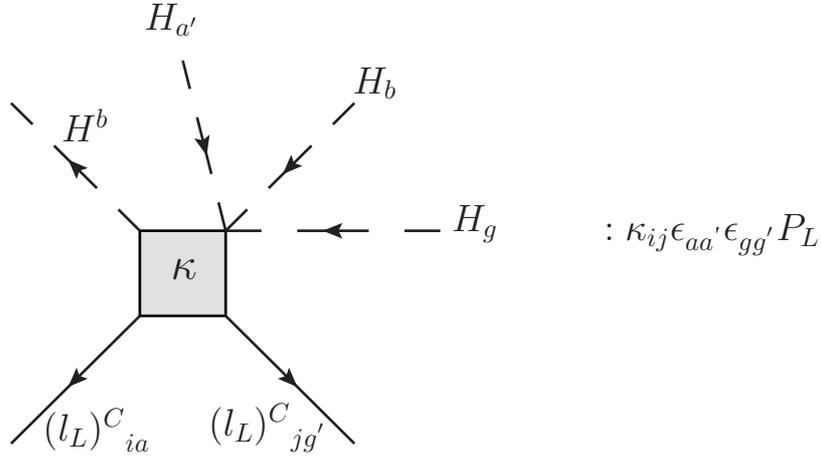}
\caption{Feynman rules for the effective vertices.}
\end{center}
\label{LLHH_effective}

\end{figure}
\end{itemize}


\end{document}